%% file: main.tex
\documentclass[aps,prd,twocolumn,showpacs,notitlepage,eqsecnum,superscriptaddress,nofootinbib]{revtex4-2}
\usepackage{comment}
\usepackage[centertags]{amsmath} \usepackage{amssymb} 
\usepackage{amsthm}
\usepackage{latexsym}
\usepackage{enumerate} \usepackage{graphicx} \usepackage{mathrsfs}
\usepackage[colorlinks]{hyperref} 
\usepackage{import}
\usepackage[usenames,dvipsnames]{xcolor} \usepackage{bm}
\usepackage{multirow} \usepackage{breakurl} \usepackage{float}
\usepackage{verbatim} \usepackage{mathrsfs}
\usepackage{booktabs}
\usepackage{orcidlink}

\usepackage{tikz}
\usetikzlibrary{arrows.meta,decorations.pathreplacing}
\usepackage{hyperref}

\newcommand{\dd}{\mathrm{d}}

\def\rcurs{{\mbox{$\resizebox{.09in}{.08in}{\includegraphics[trim= 1em 0 14em 0,clip]{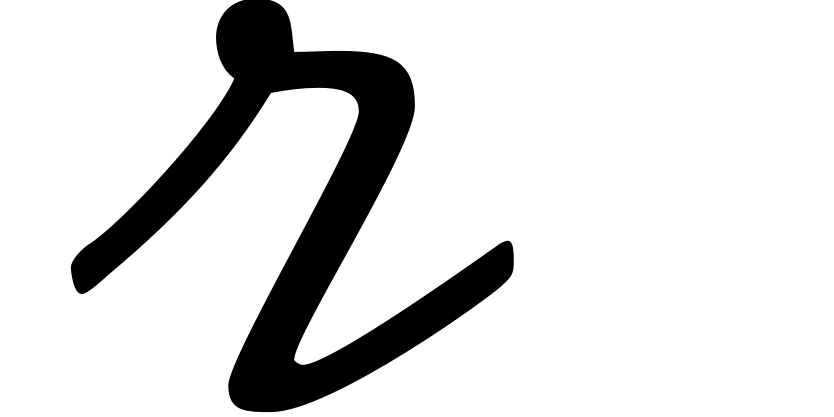}}$}}}

\allowdisplaybreaks[1]

\definecolor{CiteColor}{rgb}{0,0.5,0} \hypersetup{citecolor=CiteColor}
\definecolor{RefColor}{rgb}{0.55,0,0} \hypersetup{linkcolor=RefColor}
\definecolor{darkgreen}{rgb}{0.2,0.7,0.2}
\definecolor{red}{rgb}{1,0,0}

\begin{document}
\title{Eccentric self-force via elliptic equations with a simplified electromagnetic model}
\date{\today}

\newcommand{\geneseo}{\affiliation{Department of Physics and Astronomy, State University of New York at Geneseo, New York 14454, USA}}
\newcommand{\ucd}{\affiliation{ School of Mathematics and Statistics, University College Dublin, Belfield, Dublin 4, Ireland}}
\newcommand{\cornell}{\affiliation{Cornell Center for Astrophysics and Planetary Science, Cornell University, Ithaca, New York, 14850, USA}}
\newcommand{\ubuff}{\affiliation{Canon Stroke and Vascular Research Center, University at Buffalo, The State University of New York, 14260, USA}}

\author{Zachary L. Boice\,\orcidlink{0009-0005-3959-957X}} \email{zboice05@gmail.com} \geneseo
\author{Catherine Robb\,\orcidlink{0009-0007-1336-5307}} \email{catherinedrobb@gmail.com} \geneseo
\author{Thomas Osburn\,\orcidlink{0000-0003-2747-3994}} \email{tosburn3@gmail.com} \geneseo \ucd
\author{Joseph Balsells\,\orcidlink{0009-0000-1451-3796}} \cornell
\author{Luis R. Vazquez III\,\orcidlink{0009-0008-2167-3534}} \ubuff \geneseo

\begin{abstract}
    We investigate numerical techniques to predict gravitational waves from extreme mass-ratio inspirals with a simplified model involving self-force style calculations for eccentric motion of an electric charge in Minkowski spacetime. We solve the Lorenz gauge field equations by separating the $\phi$ and $t$ dependence, which is the frequency-domain $m$-mode method originally designed to target higher-order Kerr metric perturbations; this work implements that method with eccentric motion for the first time. We solve a set of 2D elliptic partial differential equations driven by effective sources across two puncture orders for each frequency mode and measure the frequency convergence rate for orbits with eccentricities up to 0.5. We find that non-smoothness of the effective source necessitates a large number of frequency modes beginning at moderate eccentricities. These toy-model results suggest that efficiently providing generic Kerr second-order self-force waveform templates for LISA may necessitate new techniques to achieve rapid frequency-mode convergence.
\end{abstract}

\maketitle

\section{Introduction}

The first gravitational waves were observed by the laser interferometer gravitational-wave observatory (LIGO) just over a decade ago~\cite{Abbott_2016}, and continued observations since then have offered much information about the nature of compact objects. As the LIGO/Virgo/KAGRA collaboration finishes its fourth observing run~\cite{Abac_2026}, attention has turned to the next generation of gravitational wave detectors. One such detector, the Laser Interferometer Space Antenna (LISA), is expected to launch in the 2030s and will be the first space-based gravitational wave detector~\cite{Heffernan_2026}. LISA will detect gravitational waves with far lower frequencies than are achievable with ground-based detectors (down to the millihertz range)~\cite{Babak_2021}. LISA's low-frequency sensitivity will open a window to new sources such as extreme mass-ratio inspirals (EMRIs), in which a stellar-mass compact object gradually spirals into a supermassive black hole. These systems offer a powerful probe of strong-field gravity and an opportunity to precisely test general relativity~\cite{Babak_2017}. Realizing these ambitious goals will require delivery of fast and accurate waveform models, including non-circular EMRI waveforms, before LISA begins observing. In this paper, we investigate new computational techniques for predictions involving eccentric motion in a simplified electromagnetic model as a step towards accurate non-circular EMRI waveform production.

There are three main theoretical approaches to modeling compact binary dynamics: post-Newtonian theory~\cite{Blanchet_2014}, numerical relativity~\cite{Brugmann_2018}, and black hole perturbation theory~\cite{Pound_2021} (with effective one-body theory~\cite{Buonanno_1999} synthesizing results from all three). Post-Newtonian theory deteriorates late in the inspiral when the motion becomes increasingly relativistic, and numerical relativity encounters a catastrophic separation of scales for the EMRI case; therefore, perturbation theory is the only realistic path to accurate models for relativistic EMRIs. In this approach, the dynamics are expanded in powers of the small mass ratio $\mu/M$, where $\mu$ and $M$ are the masses of the secondary and the primary. The mechanism that drives orbital decay is the gravitational self-force~\cite{Mino_1997,Quinn_1997}, which arises from the secondary's interaction with its own gravitational field~\cite{Barack_2009,Wald_2011,Poisson_2004}. Accurate self-force calculations are essential for LISA waveform models: an inspiral accumulates roughly $M/\mu \sim 10^{5}$ radians of orbital phase, and models must predict this phase with an absolute error below one radian. Meeting this accuracy requirement requires dissipative self-force calculations through second-order and conservative self-force calculations through first-order~\cite{Hinderer_2008,Pound_2021}.

To date, the problem of calculating second-order self-force effects for non-circular orbits and/or a Kerr background are unsolved. The traditional approach is to solve the field equations by full separation of variables; this Teukolsky method~\cite{Teukolsky_1972} employs spin-weighted spheroidal harmonics and produces ordinary differential equations governing each radial function. However, full separation of variables leads to problematic mode coupling~\cite{Miller_2016} and slow polar convergence when constructing the second-order source, both of which are exacerbated for Kerr. A promising strategy to ameliorate second-order source construction is to retain the $\theta$ dependence of each mode coefficient when separating into azimuthal $m$-modes. This $m$-mode approach has been explored in both the time-domain and the frequency-domain. Prior time-domain work involving $2+1$ hyperbolic partial differential equations (PDEs)~\cite{Barack_2007a,Barack_2007b,Dolan_2011a,Dolan_2011b,Thornburg_2017,Thornburg_2020} encountered unchecked growth in time~\cite{Dolan_2013}, which partly motivates this frequency-domain approach (although, work is underway to address such growth~\cite{Thornburg_2026}). Furthermore, frequency-domain calculations fit conveniently within the second-order self-force multiscale framework~\cite{Miller:2020bft}. Upon entering the frequency domain, the $m$-mode field equations become 2D elliptic PDEs. Building on recent circular motion results based on similar logic~\cite{Osburn_2022,Macedo_2024,vu2026selfforcecalculationsnumericalrelativity,osburn2026schwarzschildperturbationslorenzgauge,nishimura2026,leather2026b,cownden2026,casals2026}, this work extends the frequency-domain $m$-mode self-force approach involving elliptic PDEs to the eccentric case for the first time. 

The primary goal of this work is to use a toy model based on relativistic electrodynamics to investigate how well the frequency-domain $m$-mode strategy generalizes to eccentric motion. Specifically, we consider the electromagnetic field produced by a classical point charge in a bound eccentric orbit around a fixed opposite charge in Minkowski spacetime. In Lorenz gauge, electromagnetism and linearized gravity reduce to wave equations with distributional point-particle sources; these wave equations are qualitatively similar to the PDEs encountered in black hole perturbation theory, which is compelling for development of associated numerical techniques. Furthermore, exact solutions are available for electromagnetism, which enables immediate code validation and facilitates comprehensive convergence testing.

Applying the frequency-domain $m$-mode approach, we abstain from full separation of variables and retain the $\theta$ dependence of each mode coefficient. This produces a separate elliptic PDE for each frequency mode, which we solve numerically via finite differences. We regularize the near-field Coulomb divergence via the effective source method~\cite{Detweiler_2003,Detweiler_2003b,Vega_2008,Vega_2011,Wardell_2012,Heffernan_2012,Heffernan_2014,Bourg_2024}; these effective sources are non-smooth, and eccentric motion propagates the spatial non-smoothness at the particle into equivalent non-smoothness in time. This temporal non-smoothness inflicts an unfortunate transition from exponential frequency convergence to slower power-law convergence, which is an instance of the Gibbs phenomenon. To properly quantify generalizability to eccentric motion, we prioritize investigation of how puncture order, distance from points of non-smoothness, and eccentricity affect the frequency mode convergence of the numerical regular field.

This paper is organized as follows: Sec.~\ref{sec:theory} discusses relevant electrodynamics and self-force concepts, Sec.~\ref{sec:frequency_domain} introduces the frequency-domain $m$-mode effective source strategy, Sec.~\ref{sec:numerical_approach} shares the details of our discrete Fourier representation of the time dependence, finite-difference representation of the 2D spatial dependence, and elliptic PDE formulation as a sparse matrix problem, Sec.~\ref{sec:results} presents and analyzes our results, and Sec.~\ref{sec:conclusions} summarizes our conclusions and discusses future work. The metric has signature $(-,+,+,+)$ throughout, and electromagnetic quantities are expressed in Gaussian units with $c=1$, except where $c$ is explicitly restored.

\section{Theoretical background}
\label{sec:theory}

\begin{figure}
    \centering
    \includegraphics[width=\linewidth]{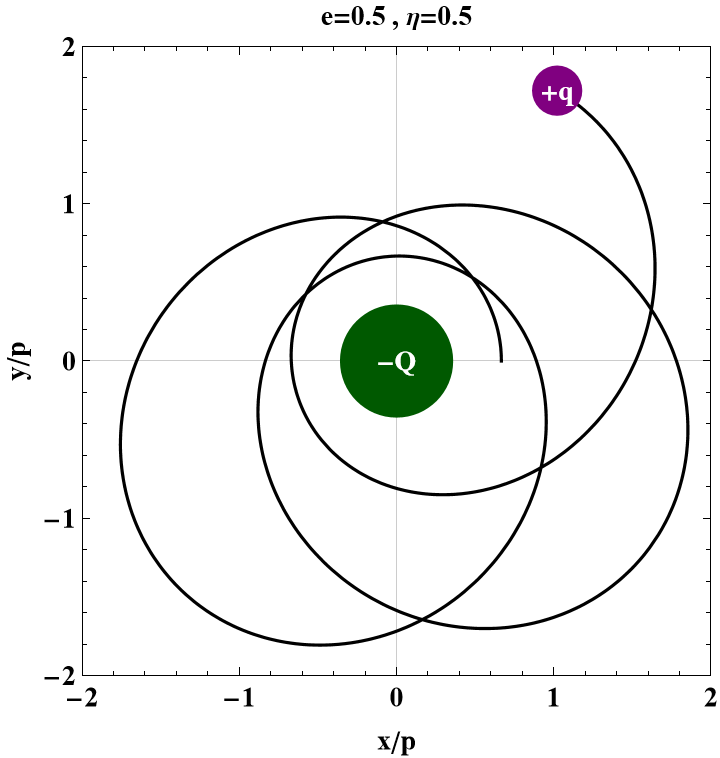}
    \caption{Motion of the smaller charge (purple) around the larger charge (green) over 2.5 radial periods. Notice how relativistic electrodynamics enables precession of pericenter, which improves the analogy with the general relativity application.}
    \label{fig:eccentric}
\end{figure}

Our model, based on classical relativistic electrodynamics, consists of a central negative point charge $-Q$ (with $Q$ defined to be positive), a smaller orbiting positive charge $q$, and the electromagnetic field tensor $F^{\alpha\beta}$, see Fig.~\ref{fig:eccentric}. As a simplifying assumption, we treat $Q$ as a non-dynamical source fixed at the origin, represented by the prescribed current density: $J_{Q}^{\alpha}=-Q \, \delta^{(3)}(\vec{x}) \, \delta^{\alpha}_{t}$, with the Dirac and Kronecker deltas respectively and where $\vec{x}$ is a spatial position vector representing the field point.

\subsection{Field equations and perturbative strategy}

Maxwell's equations govern $F^{\alpha\beta}$,
\begin{equation}
    \partial_{\beta}F^{\beta\alpha}=-4\pi J^{\alpha},
    \qquad J^{\alpha}=J_{Q}^{\alpha}+j^{\alpha},
\end{equation}
where the orbiting particle's current density is
\begin{equation}
    j^{\alpha}(x)=q\int u^{\alpha}(\tau)\,
    \delta^{(4)}\bigl(x-x_{p}(\tau)\bigr)\,\dd\tau \, ,
\end{equation}
and $u^{\alpha} = \dd x^\alpha_{p}/ \dd \tau$ is the four-velocity.
In inertial Cartesian coordinates, imposing the Lorenz gauge yields
\begin{equation}
    \Box A^{\alpha}=-4\pi J^{\alpha},
    \qquad\qquad \partial_{\alpha}A^{\alpha}=0,
    \label{eq:total_field_equation}
\end{equation}
with $\Box=\eta^{\beta\gamma}\partial_{\beta}\partial_{\gamma}$. The small charge formally obeys the relativistic Lorentz-force equation,
\begin{equation}
    \mu\frac{\dd u^{\alpha}}{\dd\tau}
    =q\, u_{\beta}\, F^{\alpha\beta}\big|_{x=x_{p}} \, ,
    \label{eq:full_particle_equation}
\end{equation}
where $\mu$ is the mass of $q$. Two obstructions prevent a straightforward solution. First, the motion and field equations are coupled. Second, Eq.~\eqref{eq:full_particle_equation} is singular for a point charge because of the local Coulomb divergence of its own field. Both of these issues are addressed  below.

By linearity of Maxwell's equations, we separate the fields sourced by the central and orbiting charges,
\begin{equation}
    A^{\alpha}=A_{Q}^{\alpha}+a^{\alpha},
    \qquad F^{\alpha\beta}=F_{Q}^{\alpha\beta}+f^{\alpha\beta},
\end{equation}
where $a^{\alpha}$ and $f^{\alpha\beta}$ are sourced by $q$ alone. Choosing each potential in Lorenz gauge gives
\begin{align}
    \Box A_{Q}^{\alpha} &=-4\pi J_{Q}^{\alpha},
    &\partial_{\alpha}A_{Q}^{\alpha}&=0, \notag\\
    \Box a^{\alpha} &=-4\pi j^{\alpha},
    &\partial_{\alpha}a^{\alpha}&=0 \, .
    \label{eq:particle_field_equation}
\end{align}

Now the particle equation takes the form
\begin{align}
    \mu\frac{\dd u^{\alpha}}{\dd\tau}
    ={}&\underbrace{q \,u_{\beta}\, F_{Q}^{\alpha\beta}}_{\propto qQ}
    +\underbrace{q\,u_{\beta}\, f^{\alpha\beta}}_{\propto q^{2}} \, .
    \label{eq:regularized_particle_equation}
\end{align}
In principle, appropriate regularization replaces the singular $f^{\alpha\beta}$ with a finite regular self-field tensor $f_{\mathrm{R}}^{\alpha\beta}$, see Sec.~\ref{sec:analytic_regular_field}. For $q\ll Q$, this force hierarchy motivates an expansion of the orbital motion in $\epsilon=q/Q$,
\begin{align}
    x_{p}^{\alpha} &= x_{0}^{\alpha}+\epsilon\, x_{1}^{\alpha}
    +\mathcal{O}(\epsilon^{2}), \notag\\
    u_{p}^{\alpha} &= u_{0}^{\alpha}+\epsilon\, u_{1}^{\alpha}
    +\mathcal{O}(\epsilon^{2}) \, .
    \label{eq:worldline_expansion}
\end{align}
Note that this is analogous to the perturbative schemes used in gravitational self-force calculations for asymmetric mass compact binaries. The leading contribution to the motion is determined by the background field alone,
\begin{equation}
    \mu\frac{\dd u_{0}^{\alpha}}{\dd\tau}
    =q\,u_{0\beta}\, F_{Q}^{\alpha\beta}\big|_{x=x_{0}} \, .
    \label{eq:background_particle_equation}
\end{equation}
Our solution strategy therefore begins by determining the background orbit (Sec.~\ref{sec:orbital_motion}), which supplies the source for the leading field of the small charge. In flat spacetime, its regular field can be obtained either analytically, by subtracting the singular part of the Li\'{e}nard--Wiechert field, or numerically, by solving effective-source equations that incorporate the regularization through a puncture as is routinely done in gravitational self-force calculations. We pursue both constructions (Secs.~\ref{sec:exact_field} and \ref{sec:frequency_domain}) and compare the resulting regular scalar potentials at the particle (Sec.~\ref{sec:results}). A full self-force treatment would then return to Eq.~\eqref{eq:regularized_particle_equation}, using the regular field tensor to determine the correction $x_{1}^{\alpha}$ to the motion\footnote{That final orbital correction is beyond the scope of this paper.}.

\subsection{Orbital motion}
\label{sec:orbital_motion}

Bound eccentric orbital motion of $q$ under the influence of $Q$ (see Fig.~\ref{fig:eccentric}) provides the current density $j^\alpha$ that produces $a^\alpha$. The covariant equation of motion is already given by Eq.~\eqref{eq:background_particle_equation}. We now express its solutions in the central charge's rest-frame coordinates. The motion is planar, and we choose the orbital plane to be $\theta=\pi/2$.

In the central charge's rest frame, the fixed background field satisfying Eq.~\eqref{eq:particle_field_equation}  is the Coulomb potential
\begin{equation}\label{AQ}
   A_{Q}^{\alpha}=\left(-\frac{Q}{r},\vec{0}\right),
    \qquad r=|\vec{x}|,
\end{equation}
with $\vec{E}_{Q}=-Q\hat{r}/r^{2}$ and $\vec{B}_{Q}=\vec{0}$.
Because we neglect $x^\alpha_1$ and $u^\alpha_1$, moving forward we adopt the notation $x^\alpha_0\rightarrow x^\alpha_p$ and $u^\alpha_0 \rightarrow u^\alpha$ for brevity. In this background potential the particle action has the proper-time Lagrangian
\begin{equation}
    L=-\mu\sqrt{-\eta_{\alpha\beta}u^{\alpha}u^{\beta}}
    +u^{t}\frac{qQ}{r_p} \, .
    \label{eq:proper_time_lagrangian}
\end{equation}
We deparametrize the motion using coordinate time $t_p(\tau)$ as the evolution parameter. An overdot denotes $\dd/\dd t_p$, so $u^{r}=u^{t}\dot{r}_p$ and $u^{\phi}=u^{t}\dot{\phi}_p$. Writing $L\,\dd\tau=L_{t}\,\dd t_p$ gives
\begin{equation}
    L_{t}=\frac{L}{u^{t}}
    =-\mu\sqrt{1-\dot{r}_p^{2}-r_p^{2}\dot{\phi}_p^{2}}
    +\frac{qQ}{r_p} \, .
    \label{eq:coordinate_time_lagrangian}
\end{equation}

Time-translation and rotational symmetry give the conserved specific energy and angular momentum,
\begin{align}
    \mathcal{E} &= \gamma-\frac{qQ}{\mu r_p},
    \label{eq:specific_energy} \\
    \mathcal{L} &= \frac{1}{\mu}\frac{\partial L_{t}}{\partial\dot{\phi}_p}
    =\gamma \, r_p^{2}\,\dot{\phi}_p,
    \label{eq:specific_angular_momentum}
\end{align}
where $\gamma=u^{t}=(1-\dot{r}_p^{2}-r_p^{2}\dot{\phi}_p^{2})^{-1/2}$. 
Solving the conserved-quantity relations for $\dot{\phi}_p$ yields
\begin{equation}\label{phidot}
\dot{\phi}_p = \frac{\mathcal{L}}{r_p^{2}}\left( \mathcal{E}+\frac{qQ}{\mu r_p} \right)^{-1} \, .
\end{equation}
Similarly, substituting $\gamma=\mathcal{E}+qQ/(\mu r_p)$ into the normalization constraint, $\eta_{\alpha\beta}\,u^{\alpha}u^{\beta}=-1$, gives
\begin{equation}\label{rdot}
\dot{r}_p^{2} = 1 - \left( 1+\frac{\mathcal{L}^{2}}{r_p^{2}} \right)\left( \mathcal{E} + \frac{qQ}{\mu r_p} \right)^{-2} \, .
\end{equation}

One convenient way of representing the time dependence of $r_p$ is through the Darwin-style radial phase $\chi(t_p)$~\cite{darwin1959gravity}
\begin{align}
r_p =\frac{p}{1+e\cos{\chi}}\, ,
\end{align}
 where $p$ is the dimensionful semi-latus rectum and $e$ is the eccentricity. Non-relativistic bound Newtonian orbits would be closed with $\chi = \phi_p$, but this is not generally true for our relativistic scenario~\cite{Minkowski1908,Sommerfeld1915}.

This change of variables  produces the equations of motion
\begin{align}
&\frac{\dd r_p}{\dd \chi} = \frac{p\,e\sin{\chi}}{(1+e\cos{\chi})^{2}} \, ,
\label{dr0 dchi}
\\
&\frac{\dd t_p}{\dd \chi} = \frac{1}{\dot{r}_p} \frac{\dd r_p}{\dd \chi} \, ,
\\
&\frac{\dd \phi_p}{\dd \chi} = \dot{\phi}_p \frac{\dd t_p}{\dd \chi} \, .
\label{dphi0 dchi}
\end{align}
When $\dot{\phi}_p$ and $\dot{r}_p$ are determined from Eqs.~\eqref{phidot} and \eqref{rdot} they refer to both the constants ($\mathcal{E}$, $\mathcal{L}$) and ($p$, $e$). These sets of constants may be related by computing the radial extrema in two ways.

On the one hand, extrema of the radial motion are accessible from $p$ and $e$ as
\begin{align}
& r_{\text{max}} = \frac{p}{1-e} \, ,
\\
& r_{\text{min}} = \frac{p}{1+e} \, .
\end{align}
These extrema are also determined by the $\dot{r}_p = 0$ roots of Eq.~\eqref{rdot}. Using these two representations we find

\begin{align}
&\mathcal{E} = - \frac{1-e^{2}}{2} \eta + \sqrt{1+\left(\frac{1-e^{2}}{2}\eta\right)^{2}} \, ,
\\
&\mathcal{L}^{2} = p^{2} \eta\left(
\frac{1+e^{2}}{2}\eta + \sqrt{1+\left(\frac{1-e^{2}}{2}\eta\right)^{2}}\right) \,,
\end{align}
where the dimensionless parameter, $\eta$, with $c$ temporarily restored is
\begin{align}
\eta \equiv \frac{qQ}{\mu p c^{2}} \, .
\end{align}

In Gaussian units, $qQ/p$ has dimensions of energy, so $\eta$ compares the Coulomb energy scale with the rest energy $\mu c^{2}$. The characteristic Newtonian orbital speed satisfies $v_{\mathrm{N}}^{2}\sim qQ/(\mu p)$, giving $\eta\sim v_{\mathrm{N}}^{2}/c^{2}$. Thus the nonrelativistic limit corresponds to $\eta\ll1$ at fixed eccentricity, and the Newtonian equations follow by retaining the leading terms in an expansion in $\eta$. All other equations use $c=1$. 

The considerations above give the remaining evolution equations
\begin{align}
& \frac{\dd t_p}{\dd \chi} = \frac{p\left(\eta(1+e^{2}+2e\cos{\chi})+\sqrt{4+\eta^{2}(1-e^{2})^{2}} \right)}{(1+e\cos{\chi})^{2}\sqrt{2\eta\left( \sqrt{4+\eta^{2}(1-e^{2})^{2}} - \eta(1-e^{2}) \right)}} \, ,
  \label{dt dchi} \\
&\frac{\dd \phi_p}{\dd \chi} = \sqrt{1+\eta\left( \eta(1-e^{2}) + \sqrt{4+\eta^{2}(1-e^{2})^{2}} \right)/\,2\;} \, .
\label{dphi dchi}
\end{align}
In particular the latter is a constant and because $\dd \phi_p/\dd \chi \ne 1$, the periastron will precess.

The period of radial oscillation, $T_{r}$, and radial angular frequency, $\Omega_{r}$, follow from integrating the $\dot{r}_p$ equation
\begin{align}
\label{eq:Tr}
&T_{r} = \frac{\pi p}{\sqrt{2\eta}} \left( \frac{\eta(1-e^{2})+\sqrt{4+\eta^{2}(1-e^{2})^{2}}}{1-e^{2}} \right)^{3/2} \, , 
\\
&\Omega_{r} = \frac{2\pi}{T_{r}} = \frac{\sqrt{\eta}}{p} \left( \frac{2(1-e^{2})}{\eta(1-e^{2})+\sqrt{4+\eta^{2}(1-e^{2})^{2}}} \right)^{3/2} \, .
\end{align}
This and the prior result  determine the average rotational frequency, $\Omega_{\phi}$,
\begin{align} \label{OmegaPhi}
& \Omega_{\phi} = \Omega_{r} \frac{\dd \phi_p}{\dd \chi} \, .
\end{align}

The first-order ODE system, Eqs.~\eqref{dr0 dchi}, \eqref{dt dchi}, and \eqref{dphi dchi}, determine the fully-relativistic background orbit $x_p^{\alpha}$ and four-velocity $u^{\alpha}$ needed to source the leading field of the orbiting charge in the chosen frame. We numerically solve the system of ODEs as an initial value problem using \texttt{NDSolve} in Mathematica.

\subsection{Local expansion of Li\'{e}nard--Wiechert solutions}
\label{sec:exact_field}
The background orbit $x_p^\alpha(\tau)$ and four-velocity $u^\alpha(\tau)$ determined in Sec.~\ref{sec:orbital_motion} specify the source according to Eq.~\eqref{eq:particle_field_equation}. In flat spacetime, the corresponding Li\'{e}nard--Wiechert potentials provide an exact benchmark for numerical solutions and a starting point for the local expansion used for regularization by constructing a puncture and effective source.

For the prescribed timelike worldline, the retarded and advanced Li\'{e}nard--Wiechert potentials take the covariant form
\begin{equation}
    a_{\pm}^{\alpha}(x)
    =\frac{q\,u^{\alpha}(\tau_{\pm})}
    {\pm u_{\beta}(\tau_{\pm})\,\gamma^{\beta}(x,\tau_{\pm})},
    \label{eq:LW}
\end{equation}
where $\gamma^{\alpha}(x,\tau)=x^{\alpha}-x_p^{\alpha}(\tau)$ and the retarded or advanced proper times, $\tau_\pm$, satisfy
\begin{equation}
\gamma_{\alpha}(x,\tau_{\pm})\,\gamma^{\alpha}(x,\tau_{\pm})=0 \, .
    \label{lightcone}
\end{equation}
The lower sign denotes the physical retarded solution, with $x_p(\tau_{-})$ on the past light cone of $x$. The upper sign denotes the advanced solution, with $x_p(\tau_{+})$ on its future light cone. Occasionally, we find it convenient to use the text ``ret"/``adv" as an equivalent way to indicate the type of solution: $a_-^{\alpha} = a_\text{ret}^\alpha$, $a_+^{\alpha} = a_\text{adv}^\alpha$.

The potentials of Eq.~\eqref{eq:LW} are exact, but their evaluation requires numerically or analytically solving the implicit light-cone condition, Eq.~\eqref{lightcone}, for $\tau_\pm$. For evaluating the Li\'{e}nard-Wiechert potentials directly we solve for $\tau_\pm$ via numerical root-finding. For a near-worldline expansion, we instead analytically solve the null condition order by order using local worldline data. Schematically, dependence on the prescribed worldline can be written as
\begin{equation}
  \tau_\pm(x)=\mathcal{F}_{\tau,\pm}[x;x_p],
  \qquad
  a_\pm^\alpha(x)=\mathcal{F}_{a,\pm}^\alpha[x;x_p].
\end{equation}

Changing the worldline changes both its null-related events and the position and four-velocity entering Eq.~\eqref{eq:LW}. For locally analytic worldlines, the functional dependence of the lightcone intersections $\tau_\pm$ can be coordinatized by a local Taylor expansion of the source. Approximating the source by its truncated Taylor expansion provides a controlled analytic approximation to the lightcone geometry and the Li\'{e}nard--Wiechert potentials.

\begin{figure}
  \centering
  \input{figures/lc-static-t0-down.tikz}
  \caption{Local light-cone geometry, illustrated for a static source with time increasing upward. The field point $x$ is null-related to the retarded event $x_p(\tau_-)$ and advanced event $x_p(\tau_+)$. The reference event $x_p(\tau_0)$ is separated from $x$ by the spacelike vector $\bar\gamma^\alpha$ and is not generally required to coincide with the coordinate time $t$ of the field point. The signed shifts are $s_\pm=\tau_\pm-\tau_0$, with $s_-<0<s_+$.}
  \label{generalExpansionPoint}
\end{figure}
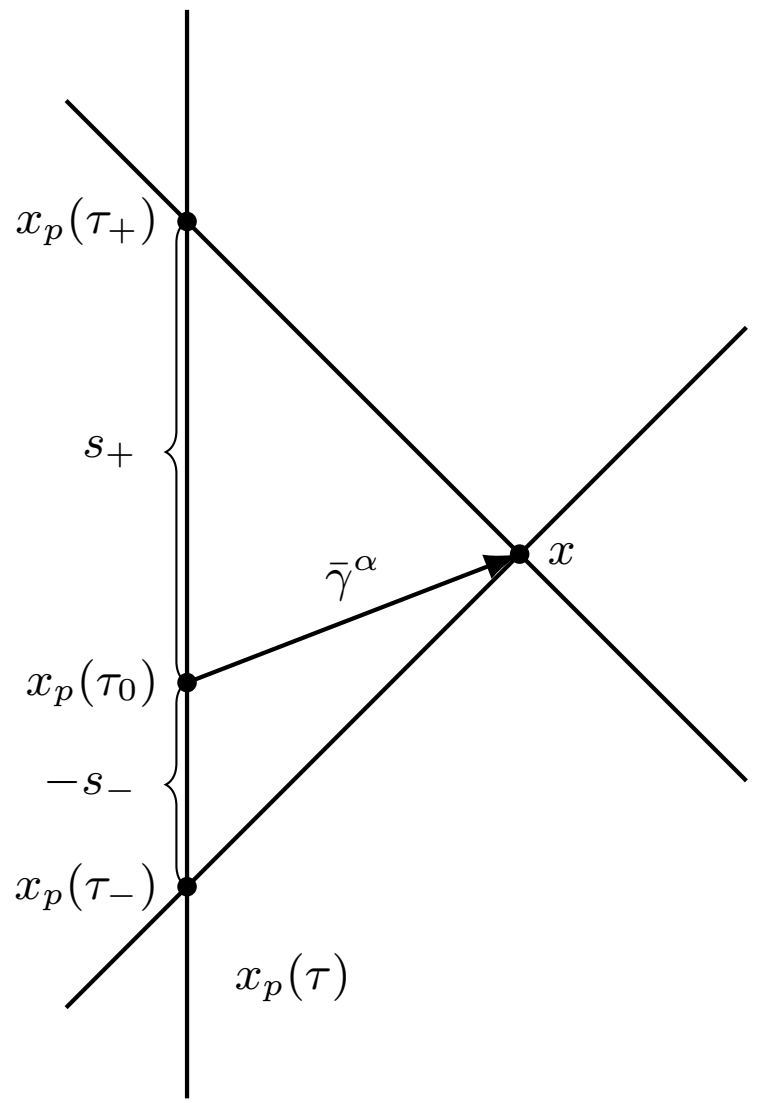

Consider a reference event on the worldline $x_p^\alpha(\tau_0)$, a nearby field point $x^\alpha$, and the related definitions:
\begin{equation}
  \bar\gamma^\alpha:=x^\alpha-x_p^\alpha(\tau_0),
  \qquad s_\pm:=\tau_\pm-\tau_0.
  \label{eq:LW_expansion_point}
\end{equation}
Evaluation at $\tau_0$ selects the expansion point on the worldline of $q$. It is convenient to assume $\bar\gamma^\alpha$ is spacelike, so that locally $\tau_-<\tau_0<\tau_+$. The reference event need not be simultaneous with $x$ in any prescribed frame, nor need $\bar\gamma^\alpha$ to be orthogonal to $u^\alpha(\tau_0)$. Figure~\ref{generalExpansionPoint} illustrates this freedom. Note that the spacetime separation vector $\bar\gamma^\alpha$ is distinct from the scalar Lorentz factor $\gamma=u^t$ used in Sec.~\ref{sec:orbital_motion}.

For a sufficiently smooth worldline, the orbital position can be locally expanded around $\tau_0$
\begin{equation}
  x_p^\alpha(\tau_0+s)
  =\sum_{n=0}^{k}\frac{s^n}{n!}\frac{\dd^nx_p^{\alpha}}{\dd \tau^n}\Big|_{\tau=\tau_0}
  +\mathcal{O}(s^{k+1}) \, ,
  \label{eq:LW_worldline_taylor}
\end{equation}
where $\dd^nx_p^{\alpha}/\dd \tau^n$ is the $n$-th derivative of $x_p^{\alpha}$ with respect to $\tau$ such that $\dd^0x_p^{\alpha}/\dd \tau^0 = x_p^{\alpha}$, $\dd^1 x_p^{\alpha}/\dd \tau^1 = u^{\alpha}$, and we introduce the proper acceleration $\alpha^\mu$
\begin{equation}
\label{eq:particle_four_acceleration}
\frac{\dd^2 x_p^{\mu}}{\dd \tau^2} = \alpha^{\mu} \, .
\end{equation}

To organize the local expansion, hold the worldline data at $\tau_0$ fixed and scale the field-source separation $\bar\gamma^\alpha\mapsto\lambda\bar\gamma^\alpha$ along a fixed spacelike direction, with $\lambda\to0^+$. The nearby null roots then obey $s_\pm=\mathcal{O}(\lambda)$.  Substituting Eq.~\eqref{eq:LW_worldline_taylor} into Eq.~\eqref{lightcone} determines the shifts $s_\pm$ order by order. Expanding the numerator and denominator of Eq.~\eqref{eq:LW} at those roots then expresses the near-worldline potentials in terms of $\bar\gamma^\alpha$ and a finite set of local derivatives. Their half-sum and half-difference give the corresponding singular and regular expansions defined below.

\subsubsection{Analytically expanded singular and regular fields}
\label{sec:analytic_regular_field}
  In flat spacetime, the singular and regular potentials are the half-sum and half-difference of the retarded and advanced solutions,
\begin{align}
    a_{S}^{\alpha}&=\frac{1}{2}\bigl(a_-^{\alpha}+a_+^{\alpha}\bigr),
    \notag\\
    a_{R}^{\alpha}&=\frac{1}{2}\bigl(a_-^{\alpha}-a_+^{\alpha}\bigr),
    \label{eq:analytic_field_split}
\end{align}
so that $a^{\alpha}=a_{S}^{\alpha}+a_{R}^{\alpha}$. This largely follows the logic of Dirac~\cite{Dirac1938} and is the flat spacetime instance of the Detweiler-Whiting prescription~\cite{Detweiler_2003}.
Solving the light-cone relation through the first two Taylor orders of the local worldline gives the half-sum expansion
\begin{equation}
  \begin{aligned}
  a_S^\alpha(x)={}&\frac{q\, u^\alpha}{\varrho}
  -q\left[
    \frac{(\alpha\cdot\bar\gamma)\,u^\alpha\bar\gamma^2}{2\varrho^3}
    +\frac{(u\cdot\bar\gamma)\,\alpha^\alpha}{\varrho}
  \right]\\
  &+\mathcal{O}(\varrho),
  \end{aligned}
  \label{AS-explicit}
\end{equation}
with the local rest-space distance
\begin{equation}
\varrho:=\sqrt{\bar\gamma^2+(u\cdot\bar\gamma)^2},
  \qquad \bar\gamma^2:=\eta_{\alpha\beta}\bar\gamma^\alpha\bar\gamma^\beta,
  \label{eq:LW_rest_distance}
\end{equation}
and with all worldline data evaluated at $\tau_0$. Scalar products~($\cdot$) of four-vectors use the Minkowski metric.

The displayed terms in Eq.~\eqref{AS-explicit} have degrees $-1$ and $0$ under $\bar\gamma^\alpha\mapsto\lambda\bar\gamma^\alpha$.  The singular potential carries the point-charge source. In fact, the degree $-1$ term is a boosted Coulomb potential that carries an entire inhomogeneous source term 
\begin{align}
  \label{CovariantSingularityApp}
  &\Box
  \left(
    \frac{qu^\mu(\tau_0)}{\varrho}
  \right)
  =
  \\&\;\;\;\;\;\;
  - 4 \pi q \int u^\mu(\tau_0)\, \delta^{(4)}\big(x-x_p(\tau_0)-(\tau-\tau_0)u(\tau_0)\big) \, {\rm d}\tau\,. \notag
\end{align}
Equation~\eqref{AS-explicit} retains the freedom to choose the reference event (rather than imposing a common $\tau_0$ choice such as $u\cdot\bar\gamma=0$~\cite{Dirac1938}).

The explicit  expansion of the half-difference cancels the inverse-distance and direction-dependent terms, leaving
\begin{equation}
  a_R^\alpha(x)=-q \, \alpha^\alpha(\tau_0)+\mathcal{O}(\varrho).
  \label{eq:regular_potential_worldline}
\end{equation}
The displayed lowest order term is pure gauge. Choosing a gauge function $\Lambda = q\, \alpha_\nu x^\nu$ produces\begin{equation}
  a_{R}^{\prime\alpha}(x)=\mathcal{O}(\varrho) \, .
  \label{eq:regular_potential_worldline-gauge}
\end{equation}

\subsubsection{Regular scalar potential on the worldline}
\label{sec:regular_scalar_worldline}

Our aim is to compute the regular field derived above numerically using an effective-source approach. We perform and validate this calculation in Sec.~\ref{sec: Comparison to Regular Field} by comparing the reconstructed regular field with its analytical value on the worldline. For a point-particle source, each retarded or advanced scalar potential, $\Phi_\pm=a_\pm^t$, together with the known orbital velocity at the corresponding light-cone event, determines the full four-vector potential through Eq.~\eqref{eq:LW}. Our comparison will therefore be between the regular scalar potential $\Phi^R$ computed by the effective-source scheme and the time component of Eq.~\eqref{eq:regular_potential_worldline}, evaluated in the inertial rest frame of the central charge.

In this frame, for each field event $x^\mu=(t,\vec x)$ we select $\tau_0$ for evaluating worldline quantities $x_p^\alpha(\tau_0)$ such that $t_p(\tau_0) = t$, where we denote 3D spatial vectors with arrows: $\vec{*}$. This produces local expansions with a reference time synchronized to the field point time and causes
\begin{equation}
    \bar\gamma^\mu=\bigl(0,\vec x-\vec x_p(t)\bigr).
\end{equation}
The worldline limit on this slice is $\vec x\to\vec x_p(t)$ at fixed $t$.

The scalar potential $\Phi=a^t$ obeys the time component of Eq.~\eqref{eq:particle_field_equation},
\begin{equation}
    \Box\Phi=-4\pi\rho,
    \qquad \Box=\nabla^{2}-\partial_{t}^{2},
    \label{field equation}
\end{equation}
where $\rho=j^t$. The singular-regular split of Eq.~\eqref{eq:analytic_field_split} directly applies to  $\Phi$ with
\begin{equation}
    \Phi=\Phi^S+\Phi^R, \qquad \Phi^{S}=\tfrac12(\Phi_-+\Phi_+).
    \label{eq:analytic_scalar_split}
\end{equation}

The regular potential, Eq.~\eqref{eq:regular_potential_worldline}, depends on the particle four-acceleration. To express its time component in terms of the orbit, write the coordinate velocity and acceleration as $\vec v_p=\dd\vec x_p/\dd t_p$ and $\vec a_p=\dd\vec v_p/\dd t_p$. With $v_p=|\vec v_p|$, $\gamma_p=(1-v_p^2)^{-1/2}$, and $\dd/\dd\tau=\gamma_p\,\dd/\dd t_p$, Eq.~\eqref{eq:particle_four_acceleration} becomes
\begin{align}
  u^\mu&=(\gamma_p,\gamma_p\,\vec v_p\,), \notag\\
  \alpha^{\mu}&=(\gamma_p^4\vec v_p\cdot\vec a_p \, , \gamma_p^2\vec a_p
    +\gamma_p^4(\vec v_p\cdot\vec a_p)\,\vec v_p\,) \, ,
  \label{eq:proper_coordinate_acceleration}
\end{align}
where the spatial dot product is Euclidean. On taking the worldline limit, the $\mathcal{O}(\varrho)$ remainder in Eq.~\eqref{eq:regular_potential_worldline} vanishes. The time component of the surviving term gives
\begin{equation}
    \Phi^{R}\!\left(t,r_p,\tfrac{\pi}{2},\phi_p\right)
    =-q\,\alpha^t=-q\,\gamma_p^4\,\vec v_p\cdot\vec a_p \, ,
    \label{analytic potential}
\end{equation}
with all orbital quantities evaluated at $t$. Calculating the regular field at the particle is somewhat analogous to calculating the Detweiler redshift~\cite{Detweiler_2008} that is often used to infer self-force effects in the gravitational case; although $\Phi^R$ is not gauge invariant, the numerical challenge of computing it resembles that of the redshift, so our analysis will focus on the regular field itself rather than the literal Abraham-Lorentz-Dirac electromagnetic self-force~\cite{Dirac1938}.

\subsubsection{Coordinate puncture construction}
\label{sec:puncture}

To construct a puncture on the same constant-$t$ slice, we expand the singular scalar potential in the spatial separation from the particle at time $t$. At fixed $\vec x$, write the displacement from the source at time $t_p$ as $\vec{\rcurs}(t_p)=\vec x-\vec x_p(t_p)$, its length as $\rcurs=|\vec{\rcurs}|$, and its direction as $\hat{\rcurs}=\vec{\rcurs}/\rcurs$. The notation involving $\vec{\rcurs}$ is adopted from  Griffiths~\cite{Griffiths2017}. The exact scalar potentials in the half-sum are the time components of Eq.~\eqref{eq:LW},
\begin{equation}
    \Phi_\pm(t,\vec x)
    =\frac{q}{\rcurs(1\pm\hat{\rcurs}\cdot\vec v_p)}
    \Bigg|_{t_p=t_\pm},
    \label{LW pot}
\end{equation}
where $t_-=t_{\mathrm{ret}}=t_p(\tau_-)$ and $t_+=t_{\mathrm{adv}}=t_p(\tau_+)$. The null condition~\eqref{lightcone} reads
\begin{equation}
    t-t_\pm=\mp\rcurs(t_\pm).
\end{equation}
At the field point the full retarded or advanced vector potential follows from $\Phi_\pm$ and the known orbital velocity,
\begin{equation}
a_\pm^\mu=\bigl(\Phi_\pm,\Phi_\pm\vec v_p(t_\pm)\bigr).
\end{equation}
Expanding the null condition about the simultaneous reference event gives
\begin{align}
    &t-t_\pm = \frac{\hat{\rcurs}\cdot\vec v_p\mp \sqrt{1-v_p^{2}+( \hat{\rcurs} \cdot \vec v_p )^{2}}}{1-v_p^{2}} \rcurs + \mathcal{O}(\rcurs^2) ,
\end{align}
where $\vec{\rcurs}$ and the orbital quantities on the right are evaluated at $t$. The next two orders involve $\vec a_p$ and $\dot{\vec a}_p$, respectively.

The equal-time separation $\bar\gamma^\mu=(0,\vec{\rcurs})$ reduces the rest-space distance in Eq.~\eqref{eq:LW_rest_distance} to
\begin{equation}
  \varrho^2=\rcurs^2+\gamma_p^2(\vec v_p\cdot\vec{\rcurs})^2.
\end{equation}
Taking the time component of Eq.~\eqref{AS-explicit} gives the equal-time singular expansion
\begin{equation}
  \begin{aligned}
  \Phi^S={}&\frac{q\gamma_p}{\varrho}
  -q\Big(
    \frac{\gamma_p^3\rcurs^2\big((\vec a_p\cdot\vec{\rcurs})
      +\gamma_p^2(\vec v_p\cdot\vec a_p)(\vec v_p\cdot\vec{\rcurs})\big)}
      {2\varrho^3}\\
  &\qquad\qquad\qquad
    +\frac{\gamma_p^5(\vec v_p\cdot\vec a_p)(\vec v_p\cdot\vec{\rcurs})}{\varrho}
  \Big)+\mathcal{O}(\rcurs).
  \end{aligned}
  \label{eq:scalar_singular_expansion}
\end{equation}

Truncating Eq.~\eqref{eq:scalar_singular_expansion} after a finite number of terms produces a puncture, $\Phi^\mathcal{P}$, that approximates $\Phi^S$
\begin{equation}
\Phi^S = \Phi^\mathcal{P}_{[j]} + \mathcal{O}(\rcurs^{j-1}) \, ,
\end{equation}
where $j$ is the puncture order. These punctures are needed for the effective-source field equations in Sec.~\ref{sec:effective_source}. Higher $j$ values produce smoother effective sources, which bring advantageous numerical and mode convergence properties.

Continuing the equal-time expansion of Eq.~\eqref{eq:scalar_singular_expansion} through one further order, we write $\Phi^{S}$ in powers of $\rcurs$,
\begin{align}
\Phi^{S} ={}& \frac{A_{[1]}(\hat{\rcurs},\vec v_p)}{\rcurs}
  + A_{[2]}(\hat{\rcurs},\vec v_p,\vec a_p) \notag\\
  &+ \rcurs\,A_{[3]}(\hat{\rcurs},\vec v_p,\vec a_p,\dot{\vec a}_p)
  + \mathcal{O}(\rcurs^{2}) \, ,
\label{phi of As}
\end{align}
where it is straightforward to identify $A_{[1]}$ from Eq.~\eqref{eq:scalar_singular_expansion}
\begin{align}
&A_{[1]}= \frac{q}{\sqrt{1-v_p^2+\big(\hat{\rcurs}\cdot\vec v_p\big)^2}}\, ,
\end{align}
and $A_{[2]}$ and $A_{[3]}$ are provided in Appendix~\ref{app:coeffs}.
This enables us to calculate punctures for the first three orders: $\Phi^\mathcal{P}_{[1]}$, $\Phi^\mathcal{P}_{[2]}$, and $\Phi^\mathcal{P}_{[3]}$.

\section{Frequency-domain $m$-mode scheme}
\label{sec:frequency_domain}
We now develop a frequency-domain $m$-mode formulation that combines separation in time and azimuth with effective-source regularization. The puncture requires only local singular-field information, so this strategy qualitatively reproduces the features of certain cutting-edge gravitational self-force calculations where closed-form solutions are unavailable. This section develops the mode equations, coordinate punctures, and Fourier transforms. The spatial discretization and numerical framework are described in Sec.~\ref{sec:numerical_approach}.

\subsection{Representation via $m$-modes}

The first step in our mode representation of $\Phi$ is to separate the $\phi$ dependence
\begin{equation}
\label{msum}
    \Phi(t,r,\theta,\phi) =\frac{1}{r} \sum_{m} \Psi_{m}(t, r,\theta)\, e^{i m \phi} ,
\end{equation}
where $\Psi_{m}(t, r,\theta)$ is an $m$-mode of $\Phi$. The inverse relation follows from orthogonality
\begin{equation}
\label{eq:monly}
\Psi_{m}(t, r,\theta) = \frac{r}{2\pi}\int_0^{2\pi} \Phi(t,r,\theta,\phi) \, e^{-im\phi} \, \dd \phi \, .
\end{equation}
Factoring out the $1/r$ scaling causes $\Psi_{m}$ to approach a constant amplitude at large radius, which is favorable numerically. Substituting Eq.~\eqref{msum} into Eq.~\eqref{field equation} produces the field equations governing each $\Psi_{m}$
\begin{align}
& \triangle_m \Psi_m = S_m \, ,
\\
\triangle_{m} = & \frac{\partial^{2} }{\partial r^{2}} + \frac{1}{r^{2}}\frac{\partial^{2} }{\partial \theta^{2}} + \frac{1}{r^{2} \tan \theta} \frac{\partial }{\partial \theta} - \frac{m^{2}}{r^{2}\sin^{2}\theta} - \frac{\partial^2}{\partial t^2} \, ,
\end{align}
where $S_m$ satisfies a similar orthogonality relation
\begin{equation}
\label{eq:Sm}
S_m(t, r, \theta) = \frac{r}{2\pi} \int_0^{2\pi} -4\pi\rho(t,r,\theta,\phi) \, e^{-im\phi}\, \dd\phi \, .
\end{equation}
Adopting such a time-domain picture universally can have advantages (see Refs.~\cite{Barack_2007a,Barack_2007b,Dolan_2011a,Dolan_2011b}), but would also introduce certain obstacles~\cite{Dolan_2013}. Here we use the time-domain as an intermediate step before entering the frequency domain for numerical calculations.

\subsection{Eccentric $m$-mode punctures}
It is convenient to identify a version of $\Phi^\mathcal{P}_{[1]}$ that is useful as a reference to inform higher-order punctures. One valid approximation for $\Phi^\mathcal{P}_{[1]}$ is
\begin{align}
\label{eq:phiP1}
&\Phi^{\mathcal{P}}_{[1]} = \frac{q}{R} \, ,
\\
&R^{2} = \bar{\varrho}^{2} + z_{c}^{2} \, \text{sin}^{2}\bigg(\frac{\Delta\phi}{2}  + \frac{\dot{r}_p \dot{\phi}_p \Delta r}{2(1-\dot{r}_p^{2})} \bigg) \, ,
\\
& \bar{\varrho}^{2} =r_p^{2} (1 - \dot{r}_p^{2} - r_p^{2} \dot{\phi}_p^{2}) \Delta\theta^{2} + \bigg(1 - \frac{r_p^{2} \dot{\phi}_p^{2}}{1 - \dot{r}_p^{2}} \bigg) \Delta r^{2} \, ,
\\
& z_{c}^{2} = 4 r_p^{2} (1 - \dot{r}_p^{2}) \, ,
\end{align}
where $\Delta r = r-r_p$, $\Delta \theta = \theta-\pi/2$, $\Delta \phi = \phi-\phi_p$. Notice that Eq.~\eqref{eq:phiP1} involves some ambiguity about whether the $\Delta r$, $\Delta \theta$, and $\Delta \phi$ dependence in the denominator is represented as a pure power series or with other functions (see the $\sin^2$ term); we use this ambiguity to design a version of $\Phi^{\mathcal{P}}_{[1]}$ that enables easier application of Eq.~\eqref{eq:monly}, which is allowed so long as $\Phi^S=\Phi^{\mathcal{P}}_{[1]}+\mathcal{O}(\rcurs^0)$.

The next step is integration over $\Delta\phi$ according to Eq.~\eqref{eq:monly} to calculate $\Psi^\mathcal{P}_{[j]m}$ from $\Phi^\mathcal{P}_{j}$, which suggests a change of variable to $\Delta\tilde{\phi}$ is in order
\begin{align}
\Delta\phi & = \Delta\tilde{\phi}  - \frac{\dot{r}_p \dot{\phi}_p}{1-\dot{r}_p^{2}}\Delta r \, ,
\\
R^{2} & = \bar{\varrho}^{2} + z_{c}^{2} \, \text{sin}^{2}\frac{\Delta\tilde{\phi}}{2} \, .
\end{align}
This simplification of $R$ in terms of $\Delta\tilde{\phi}$ enables straightforward analytical expressions for each $\Psi^\mathcal{P}_{[j]m}$ after integration
\begin{align}
\Psi^\mathcal{P}_{[j]m} &= \frac{r}{2\pi} \int_{0}^{2 \pi} \Phi^\mathcal{P}_{[j]}\; e^{-im\phi} \dd \phi 
\\
&= \frac{r\,e^{-im\phi_p}}{2\pi} \int_{0}^{2 \pi} \Phi^\mathcal{P}_{[j]}\; e^{-im\Delta\phi} \dd \Delta\phi \notag 
\\
&= \frac{r\,e^{-im\phi_p}}{2\pi} \int_{0}^{2 \pi} \Phi^\mathcal{P}_{[j]}\; e^{-im(\Delta\tilde{\phi} - \dot{r}_p \dot{\phi}_p\Delta r/(1-\dot{r}_p^{2}))} \dd \Delta\tilde\phi \notag
\\
&= \frac{r\,e^{-im(\phi_p-\dot{r}_p \dot{\phi}_p\Delta r/(1-\dot{r}_p^{2}))}}{2\pi} \int_{0}^{2 \pi} \Phi^\mathcal{P}_{[j]}\; e^{-im\Delta\tilde{\phi}} \dd \Delta\tilde\phi \, . \notag
\end{align}
Our initial puncture strategy is to use powers of $R$ in the denominator of $\Phi^\mathcal{P}_{[j]}$ to describe the singularity, and then re-expand the numerator as a power series in $\Delta r$, $\Delta \theta$, and $\Delta\tilde\phi$. The one difficulty is $\Delta\tilde\phi$, which should involve periodic functional dependence so that the $m$-mode integrands are smooth functions of $\Delta\tilde\phi$. This motivates us to express each power of $\Delta\tilde\phi$ in the numerator equivalently as complex exponentials
\begin{align}
\Delta\tilde\phi = \frac{i}{2} e^{-i\Delta\tilde\phi} - 
 \frac{i}{2} e^{i\Delta\tilde\phi} + \mathcal{O}(\Delta\tilde\phi^{3}) \, ,
\end{align}
where higher powers of $\Delta\tilde\phi$ accurate through higher orders are similarly available. This enables the following representation of each puncture
\begin{align}
\Phi^{\mathcal{P}}_{[j]} = \frac{1}{R^{2j-1}} \sum_{k=-j}^{j} B_{j}^{k}(t, \Delta r, \Delta\theta) \, e^{ik\Delta\tilde\phi} \, ,
\end{align}
where re-expanding the numerator of the singular field to calculate each $B_{j}^{k}$ is key. 
This is convenient because of the following relationship to calculate the integrals analytically~\cite{Bourg_2024}
\begin{align}
&\int_{0}^{2\pi} \frac{e^{-i m \Delta\tilde\phi}}{R^{2j-1}} \dd \Delta\tilde\phi = 
\\&\qquad\qquad\qquad \frac{2 \pi\,(-1)^{m} }{(3/2-j)_{m}} \Big(\bar\varrho \sqrt{\bar\varrho^{2} + z_{c}^{2}}\Big)^{1/2-j} {}_{3} P_{1/2-j}^{m}(z) \, , \notag
\\
& z = \frac{\bar\varrho^{2}+z_{c}^{2}/2}{\bar\varrho\sqrt{\bar\varrho^{2}+z_{c}^{2}}} \, ,
\end{align}
where $(x)_{m}$ is the Pochhammer symbol applied to $x$ with $m$ factors, and ${}_{3} P_{n}^{m}(z)$ is the associated Legendre function of type 3, which is defined for $z>1$. This enables a relatively straightforward expression for the $m$-modes of each puncture
\begin{align}
&\Psi_{[j]m}^{\mathcal{P}} = r\,e^{-im(\phi_p-\dot{r}_p \dot{\phi}_p \Delta r/(1-\dot{r}_p^{2}))} \times
\\&\;\;\;\;\;\;\;\; \sum_{k=-j}^{j} \frac{(-1)^{m-k} B_{j}^{k}}{(3/2-j)_{m-k}} \Big(\bar\varrho \sqrt{\bar\varrho^{2} + z_{c}^{2}}\Big)^{1/2-j} {}_{3} P_{1/2-j}^{m-k}(z) \, .\notag
\end{align}

\subsection{Effective-source regularization}
\label{sec:effective_source}
\begin{figure}
\centering
\includegraphics[trim={0 0 0 0},clip,width=\linewidth]{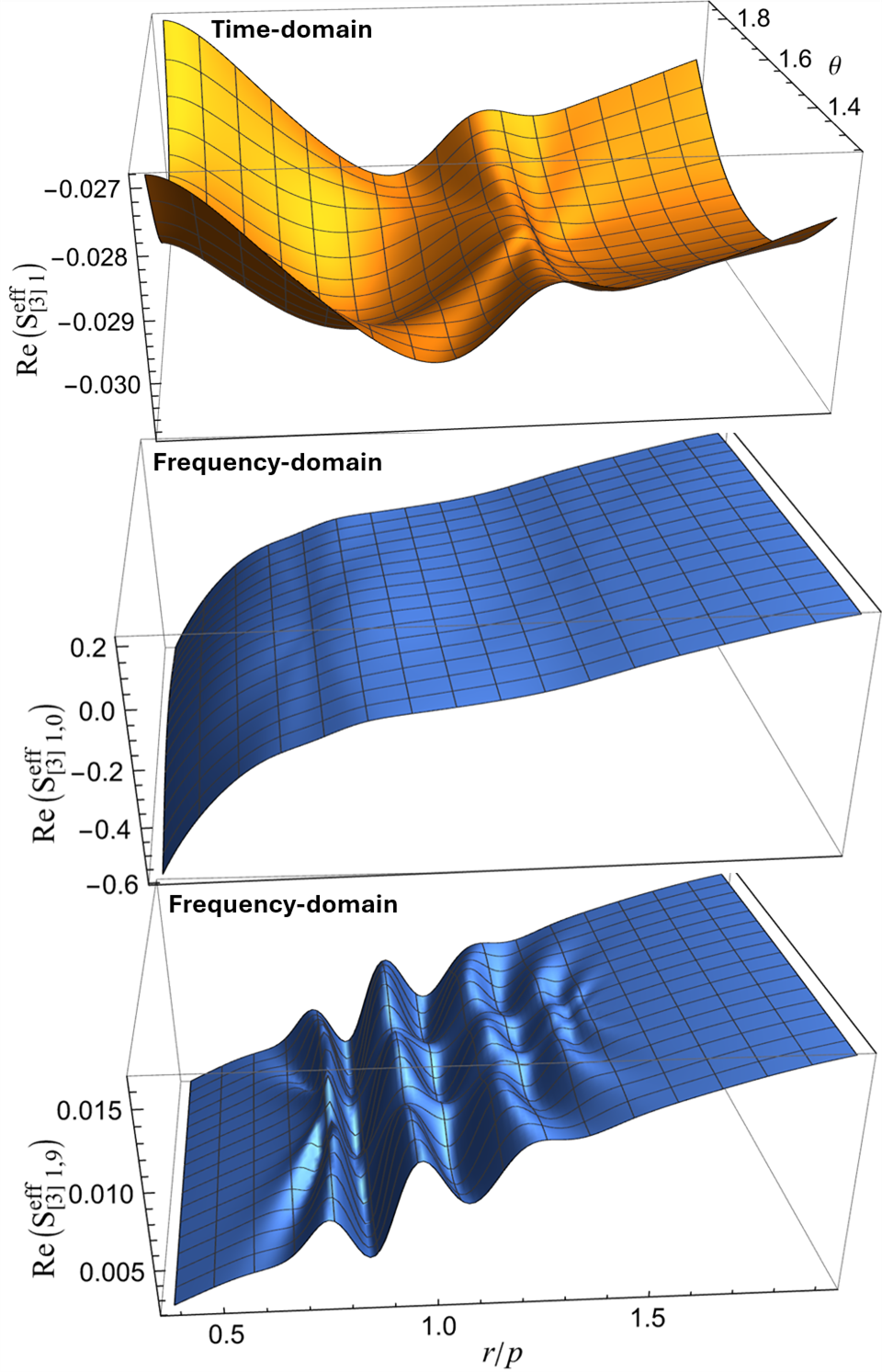}
\caption{\textit{Top}: $S_{[j]m}^{\mathrm{eff}}$ is shown for $j=3$, $m=1$, $e=0.3$, and $\eta=0.5$ at time $t/p \simeq 9.117$; notice the non-smooth ``pinch'' feature at the charge position. \textit{Middle}: the $n=0$ Fourier mode ($S_{[j]mn}^{\mathrm{eff}}$) is shown; a faint imprint of the non-smoothness is visible at a position consistent with times when the effective source is largest. \textit{Bottom}: the $n=9$ Fourier mode ($S_{[j]mn}^{\mathrm{eff}}$) is shown; higher harmonics involve small but strongly oscillatory spatial features.
}
\label{fig:Seff3D}
\end{figure}

Although the analytical split in Eq.~\eqref{eq:analytic_scalar_split} identifies the correct regular field, we seek an equivalent numerical formulation that is analogous to modern gravitational self-force calculations. Using $\Phi^{\mathcal{P}}_{[j]}$ to approximate $\Phi^{S}$ produces the residual field $\Phi^{\mathcal{R}}_{[j]}$
\begin{equation}
    \Phi^\text{ret}=\Phi_{[j]}^{\mathcal{P}}+\Phi_{[j]}^{\mathcal{R}},
    \label{eq:Phi_field_decomposition}
\end{equation}
In spacetime, the corresponding residual satisfies $\Phi^{\mathcal{R}}_{[j]}=\Phi^{R}+(\Phi^{S}-\Phi^{\mathcal{P}}_{[j]})$. It therefore agrees with the exact regular potential on the worldline if the puncture remainder vanishes there; away from the worldline $\Phi^{\mathcal{R}}$ and $\Phi^{R}$ generally differ. 
The same logic applied to $\Psi_m$ produces the $m$-modes of the residual field, $\Psi_{[j]m}^{\mathcal{R}}$
\begin{equation}
    \Psi^\text{ret}_{m}=\Psi_{[j]m}^{\mathcal{P}}+\Psi_{[j]m}^{\mathcal{R}}\, .
    \label{eq:field_decomposition}
\end{equation}
Substituting this decomposition into the field equation and rearranging gives
\begin{equation}
    \triangle_{m}\Psi_{[j]m}^{\mathcal{R}}=S_{m}-\triangle_{m}\Psi_{[j]m}^{\mathcal{P}}\equiv
    S_{[j]m}^{\mathrm{eff}}.
    \label{eq:effective_source}
\end{equation}
For a sufficiently high-order puncture, $j>2$, the effective source $S_{[j]m}^{\mathrm{eff}}$ is finite near the particle (although $j=2$ provides an infinite $S_{[2]m}^{\mathrm{eff}}$ that is nevertheless integrable). These features enable the numerical solve to incorporate an analytic expression for $S_{[j]m}^{\mathrm{eff}}$ rather than attempting to subtract two divergent numerical fields. The $j=3$ effective source is depicted in the top panel of Fig.~\ref{fig:Seff3D} for the $m=1$ case.

\subsection{Transforming to the frequency-domain}
\label{sec:frequency_decomposition}
For bound eccentric motion, the orbital radius satisfies $r_p(t+T_{r})=r_p(t)$, with $T_r$ given by Eq.~\eqref{eq:Tr}, while the azimuthal angle advances by $\phi_p(t+T_{r})-\phi_p(t)=\Omega_{\phi} T_{r}$, with $\Omega_{\phi}$ given by Eq.~\eqref{OmegaPhi}. These features imply a discrete frequency spectrum, which we represent as a Fourier series with angular frequency  $\omega_{mn}=m\Omega_{\phi}+n\Omega_{r}$
\begin{align}
&\Psi_{m}(t,r,\theta) =\sum_{n} \Psi_{mn}(r,\theta)\, e^{-i \omega_{mn}t}\, .
    \label{decomposition}
\end{align}
This reduces the numerical problem to solving field equations for each $\Psi_{mn}$ over the two spatial coordinates $(r,\theta)$.
The field equations and inverse relation follow from the same logic as for the $m$-mode case
\begin{align}
& \triangle_{mn} \Psi_{mn} = S_{mn} \, ,
\label{eq:modes DE}
\\
& \triangle_{mn} = \frac{\partial^{2} }{\partial r^{2}} + \frac{1}{r^{2}}\frac{\partial^{2} }{\partial \theta^{2}} + \frac{1}{r^{2} \tan \theta} \frac{\partial }{\partial \theta} - \frac{m^{2}}{r^{2}\sin^{2}\theta} + \omega_{mn}^2 \, ,
\label{eq:delta_mn}
\\
\label{eq:PsiFD}
&\Psi_{mn}(r,\theta) = \frac{1}{T_r}\int_0^{T_r} \Psi_{m}(t, r,\theta) \, e^{i\omega_{mn}t} \, \dd t \, ,
\\
\label{eq:SFD}
&S_{mn}(r,\theta) = \frac{1}{T_r}\int_0^{T_r} S_{m}(t, r,\theta) \, e^{i\omega_{mn}t} \, \dd t \, .
\end{align}

The domain is divided into an interior worldtube surrounding the particle and an exterior region. The unknown numerical fields are defined as the residual field $\Psi_{[j]mn}^{\mathcal{R}}$ within the worldtube and the retarded field $\Psi^\text{ret}_{mn}$ outside the worldtube. The two field representations are related at the worldtube boundary by 
\begin{equation}
    \Psi_{[j]mn}^{\mathcal{R}}=\Psi^\text{ret}_{mn}-\Psi_{[j]mn}^{\mathcal{P}}.
    \label{eq:worldtube_relation}
\end{equation}
or equivalently,
\begin{equation}
    \Psi^\text{ret}_{mn}=\Psi_{[j]mn}^{\mathcal{R}}+\Psi_{[j]mn}^{\mathcal{P}}.
    \label{eq:puncture_conversion}
\end{equation}
This transformation allows the same numerical grid to represent either the residual or retarded field while maintaining the appropriate field equation in each region. The resulting PDE system is
\begin{align}
& \triangle_{mn}\, \Psi^{\mathcal{R}}_{[j]mn} = S^{\text{eff}}_{[j]mn} \qquad\;\;\;\; (\text{interior}) \, , 
\\
& \triangle_{mn}\, \Psi^\text{ret}_{mn} = 0 \qquad\qquad\qquad (\text{exterior}) \, , 
\\
&\Psi^{\mathcal{R}}_{[j]mn} = \Psi^\text{ret}_{mn}-\Psi^{\mathcal{P}}_{[j]mn} \qquad (\text{across}) \, ,
\label{PDE system}
\end{align}
where
\begin{align}
S^{\text{eff}}_{[j]mn}(r,\theta) &=\frac{1}{T_r}\int_0^{T_r} S^{\text{eff}}_{[j]m}(t,r,\theta) \, e^{i\omega_{mn} t} \, \dd t
\\
&=\frac{1}{T_r}\int_0^{T_r} \big( S_{m} - \triangle_{m}\Psi_{[j]m}^{\mathcal{P}} \big) \, e^{i\omega_{mn} t} \, \dd t \, . \notag
\end{align}
Two $n$-modes of the $j=3$ effective source are depicted in the middle and bottom panels of Fig.~\ref{fig:Seff3D} for the $m=1$ case.

\section{Numerical Approach}
\label{sec:numerical_approach}
We now discretize in time and space to numerically solve the regularized $mn$-mode equations developed in Sec.~\ref{sec:frequency_domain}. Much of our general approach and spatial domain discretization follow Osburn and Nishimura~\cite{Osburn_2022}.

\subsection{Discrete Fourier transforms}

For each $m$, we evaluate the analytical puncture and effective source as functions of time and then compute their $n$ modes by Fourier projection,
\begin{equation}
    X_{mn}(r,\theta) = \frac{1}{T_{r}} \int_{0}^{T_{r}} \left( X_{m}(t,r,\theta) \, e^{im \Omega_{\phi} t} \right)e^{in \Omega_{r} t} \, \dd t \, ,
\end{equation}
where $X$ refers either to the effective source or puncture depending on whether Eq.~\eqref{eq:PsiFD} or Eq.~\eqref{eq:SFD} is considered. Importantly, the factor in the parentheses is periodic in time with period $T_{r}$. We perform this integral for $\Psi^\mathcal{P}_{[j]mn}$ to implement each set of jumps across the worldtube and for $S^\text{eff}_{[j]mn}$ inside the worldtube as a source to drive each $\Psi^\mathcal{R}_{[j]mn}$.

The set of Fourier integrals is rapidly evaluated using discrete Fourier transforms by sampling the puncture and effective source at $N$ equally spaced times over one radial period (including $t=0$ and excluding $t=T_{r}$). Thus, our discrete times are $t_{k} = k \, T_{r}/N$, and the following provides the discrete Fourier transform
\begin{equation}
    X_{mn}(r,\theta) \simeq \frac{1}{N} \sum_{k=0}^{N-1} \left( X_{m}(t_{k},r,\theta)\,e^{i m \Omega_{\phi} t_{k}} \right) e^{in \Omega_{r} t_{k}} \, .
\end{equation}
Applying a fast Fourier transform algorithm (FFT) to the factor in the parentheses provides all of the Fourier coefficients $X_{mn}$ (for all $n$) in one efficient calculation~\cite{Hopper_2015}. 

\subsection{Finite difference discretization}
\begin{figure}
    \centering
    \includegraphics[width=\linewidth]{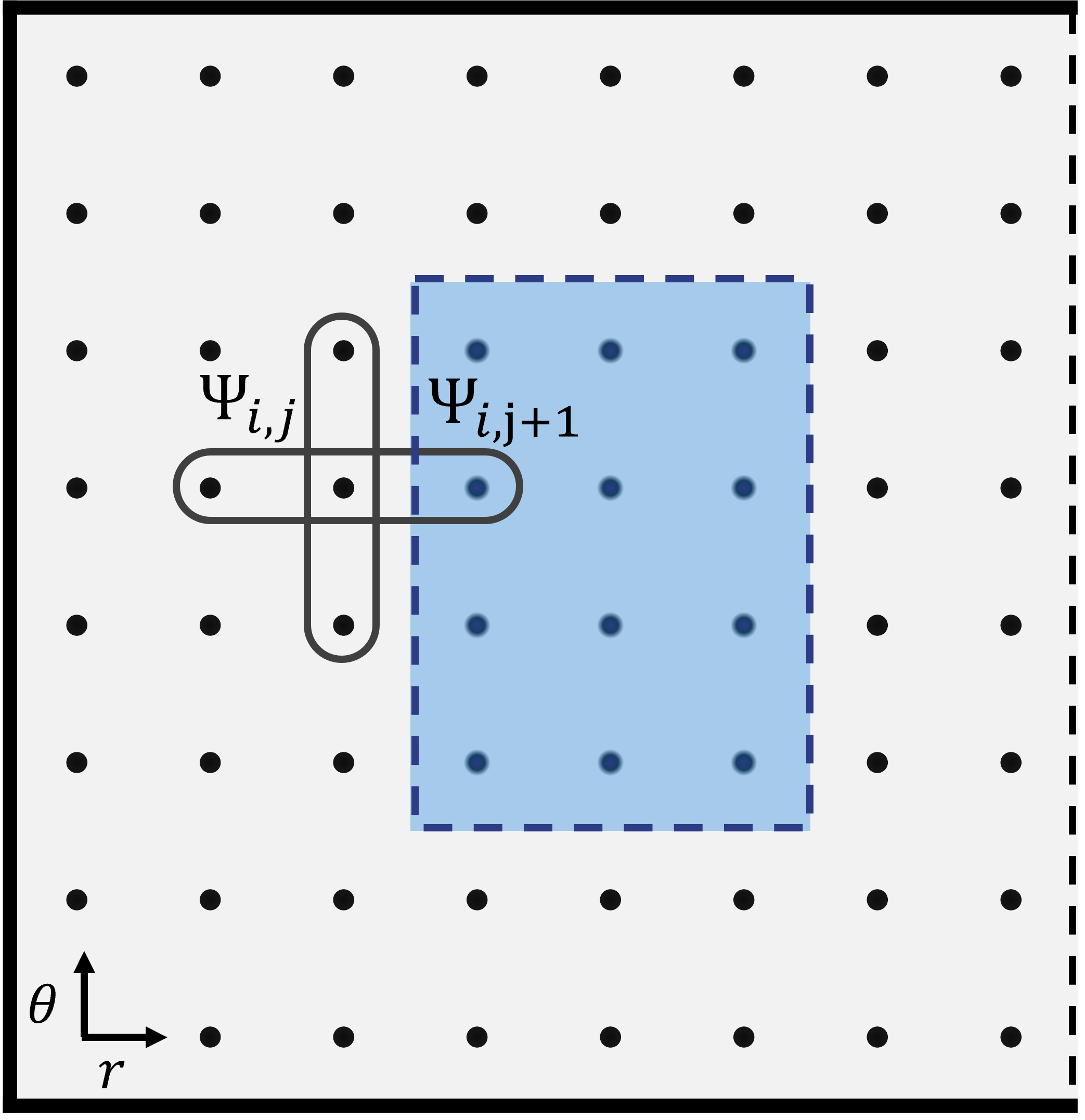}
    \caption{The discretization of the spatial domain is illustrated along with our five-point finite difference stencil. Where the stencil intersects the worldtube, we transform from $\Psi_{i,j+1}$ to $\Psi_{i,j+1}^\mathcal{R}$ using the puncture $\Psi_{i,j+1}^\mathcal{P}$. Note that our actual implementation of the worldtube involves a polar width of $\pi/5$; this low-resolution depiction is only for illustrative purposes.}
    \label{fig:worldtube}
\end{figure}

The spatial computational domain is discretized on a uniform rectangular grid. Polar and radial grid points are denoted by 
\begin{equation}
    \theta_{i}=i\Delta\theta, \qquad\qquad r_{j}=j\Delta r,
\end{equation}
where
\begin{equation}
    i=0,\ldots,N_{\theta}-1,\qquad
    j=0,\ldots,N_{r}-1,
\end{equation}
and $\Delta r$ and $\Delta \theta$ are appropriate grid spacings; note that, moving forward, we have redefined $\Delta\theta$ and $\Delta r$ where previously they were used as relative positions in the puncture expansion.  The worldtube must always contain the position of the particle: $(r_p(t), \pi/2)$. Given an eccentricity and semi-latus rectum, we can find $r_{\text{min}}$ and $r_{\text{max}}$ as the minimum and maximum radial positions of the particle. In the $r$ direction, we divide our domain into units $L_{r}=r_{\text{min}}/2$ and place the inner worldtube transition at $r=L_r$; then the worldtube extends to an outer transition at $r=l\,L_r$, where $l$ is an integer such that $(l-1)\,L_r < r_\text{max}+L_r \le l\,L_r$ (note that this condition unambiguously determines $l$). The radial domain ends at $r=b\,L_r$ where $b$ is an integer greater than $l$ that we use to control the extent of the radial domain. We define the equivalent $\theta$ unit to be the polar worldtube width of $L_{\theta}=\pi/5$, and we center the worldtube at $\theta=\pi/2$. To avoid the non-smoothness of the effective source as much as possible, we require an odd number of $\Delta\theta$ spaces to span $L_\theta$: $\Delta\theta=L_\theta/(2J_\theta+1)$, where $J_\theta$ is an integer we use to control $\theta$ resolution. We do not need to impose the odd restriction on $L_r$: $\Delta r=L_r/J_r$, where $J_r$ is an integer we use to control $r$ resolution. Through this logic, the total number of radial points is $N_r=1+b J_r$ and the total number of polar points is $N_\theta=1+5(2J_\theta+1)$. This spatial discretization is illustrated qualitatively in Fig.~\ref{fig:worldtube}.

For the remainder of this section, we drop the subscripts $mn$ on $\Psi$ and $\omega$ such that all $\Psi, \omega$ refer equally well to an arbitrary mode of the solution (cases where this is not true are stated explicitly). The derivatives appearing in Eq.~\eqref{eq:delta_mn} are replaced by second-order centered finite differences at interior grid points. The radial second derivative is approximated by
\begin{equation}
    \left.\frac{\partial^{2}\Psi}{\partial r^{2}}\right|_{ij}\approx\frac{\Psi_{i,j+1}-2\Psi_{ij}+\Psi_{i,j-1}}{\Delta r^{2}},
    \label{eq:radial_second}
\end{equation}
while the angular derivatives are approximated by
\begin{align}
&    \left.\frac{\partial\Psi}{\partial\theta}\right|_{ij}\approx\frac{\Psi_{i+1,j}-\Psi_{i-1,j}}{2\Delta\theta},
    \label{eq:theta_first}
\\
&    \left.\frac{\partial^{2}\Psi}{\partial\theta^{2}}\right|_{ij}\approx\frac{\Psi_{i+1,j}-2\Psi_{ij}+\Psi_{i-1,j}}{\Delta\theta^{2}}.
    \label{eq:theta_second}
\end{align}
Substituting Eqs.~\eqref{eq:radial_second}--\eqref{eq:theta_second} into Eq.~\eqref{eq:modes DE} gives a discrete equation of the form
\begin{equation}
d_{P}\Psi_{ij}+d_{E}\Psi_{i,j+1}+d_{W}\Psi_{i,j-1}+d_{N}\Psi_{i+1,j}+d_{S}\Psi_{i-1,j}=S_{ij},
    \label{eq:five_point}
\end{equation}
where
\begin{align}
&    d_{E}=d_{W}=\frac{1}{\Delta r^{2}},
    \label{eq:aE_aW}
\\
&    d_{N}=\frac{1}{r_{j}^{2}\Delta\theta^{2}}+\frac{1}{2r_{j}^{2}\tan\theta_{i}\,\Delta\theta},
    \label{eq:aN}
\\
&    d_{S}=\frac{1}{r_{j}^{2}\Delta\theta^{2}}-\frac{1}{2r_{j}^{2}\tan\theta_{i}\,\Delta\theta},
    \label{eq:aS}
\\
&    d_{P}=-\frac{2}{\Delta r^{2}}-\frac{2}{r_{j}^{2}\Delta\theta^{2}}-\frac{m^{2}}{r_{j}^{2}\sin^{2}\theta_{i}}+\omega^{2} ,
    \label{eq:aP}
\end{align}
and $S_{ij}$ may be zero or non-zero depending on whether ($i$,$j$) is outside or inside the worldtube. 
Thus, at an interior grid point, the finite-difference equation couples the field value only to its four nearest neighbors, producing a five-point finite-difference stencil.

To convert the two-dimensional grid of $\Psi_{ij}$'s into a one-dimensional vector $\mathbf{\Psi}$, the following composite index is used
\begin{equation}
    k=j+N_{r} i.
    \label{eq:composite_index}
\end{equation}

The radial neighbors of a point $(i,j)$ therefore correspond to 
\begin{equation}
    (i,j+1)\longrightarrow k+1,\qquad(i,j-1)\longrightarrow k-1,
\end{equation}
while the angular neighbors correspond to
\begin{equation}
    (i+1,j)\longrightarrow k+N_{r},\qquad(i-1,j)\longrightarrow k-N_{r}.
\end{equation}
We collect the coefficients of $(\mathbf{\Psi})_{k}$ for the finite difference representation of \eqref{eq:modes DE} at all grid points and represent the $N_{r}N_{\theta}\times N_{r}N_{\theta}$ coefficient matrix by $\mathbf{A}$. Then, for each $mn$ mode, we solve the sparse matrix equation
\begin{equation}
    \mathbf{A}\boldsymbol{\Psi}=\boldsymbol{S},
    \label{eq:matrix_equation}
\end{equation}
where $\boldsymbol{S}$ contains the discretized source terms.

\subsection{Worldtube and puncture treatment}
The worldtube treatment determines which field representation and source are used at each grid point. Following the worldtube and finite difference method of~\cite{Osburn_2022}, there are three cases: points entirely inside the worldtube, points entirely outside the worldtube, and finite-difference stencils that cross the worldtube boundary.
Inside the worldtube, the unknown is the regular field $\Psi^{\mathcal{R}}$, and the finite-difference equation is 
\begin{equation}
d_{P}\Psi^\mathcal{R}_{ij}+d_{E}\Psi^\mathcal{R}_{i,j+1}+d_{W}\Psi^\mathcal{R}_{i,j-1}+d_{N}\Psi^\mathcal{R}_{i+1,j}+d_{S}\Psi^\mathcal{R}_{i-1,j}=S^\text{eff}_{ij}.
\end{equation}
Outside the worldtube, the unknown is the retarded field $\Psi^\text{ret}$, and the corresponding equation is homogeneous,
\begin{equation}
d_{P}\Psi^\text{ret}_{ij}+d_{E}\Psi^\text{ret}_{i,j+1}+d_{W}\Psi^\text{ret}_{i,j-1}+d_{N}\Psi^\text{ret}_{i+1,j}+d_{S}\Psi^\text{ret}_{i-1,j}=0\, ,
\end{equation}
When a finite-difference stencil crosses the worldtube boundary, some grid values correspond to the regular field, and others correspond to the full field. The required conversion is made using 
\begin{equation}
    \Psi^\text{ret}=\Psi^{\mathcal{R}}+\Psi^{\mathcal{P}}.
\end{equation}
Consider as an example the case where the positions $(i,j)$, $(i-1,j)$, $(i+1,j)$, and $(i,j-1)$ are outside the worldtube while the position $(i,j+1)$ is inside the worldtube, see Figure~\ref{fig:worldtube}. The correct equation centered on $(i,j)$ should involve $\Psi^\text{ret}$ throughout, but the hypothetical field $\Psi^\text{ret}_{i,j+1}$ does not appear in the vector of unknowns when that position is inside the worldtube. For this to be successful, we must transform the equation to only involve unknowns present in $\mathbf{\Psi}$
\begin{equation}
d_{P}\Psi^\text{ret}_{ij}+d_{E}(\Psi^{\mathcal{R}}_{i,j+1}+\Psi^{\mathcal{P}}_{i,j+1})+d_{W}\Psi^\text{ret}_{i,j-1}+\ldots=0\, .
\end{equation}
Because $\Psi^{\mathcal{P}}_{i,j+1}$ is known prior to the solve, we move it to the right hand side and treat it as a source
\begin{equation}
d_{P}\Psi^\text{ret}_{ij}+d_{E}\Psi^{\mathcal{R}}_{i,j+1}+d_{W}\Psi^\text{ret}_{i,j-1}+\ldots=-d_E\,\Psi^{\mathcal{P}}_{i,j+1}\, .
\end{equation}
We implement this logic at all positions where the finite-difference stencil intersects the worldtube; therefore, our strategy is to appropriately insert $\Psi^{\mathcal{P}}$ into $\mathbf{S}$ to enforce the jumps across the worldtube, which does not affect the sparse matrix $\mathbf{A}$~\cite{Osburn_2022}.

\subsection{Boundary conditions}

The boundary rows of the matrix are constructed separately from the interior finite difference equations.
At the inner radial boundary, $r=0$, the condition $\Psi |_{r=0}=0$ is imposed by replacing the corresponding rows of the matrix with identity rows; note that this permits a finite $\Phi$ at the origin because $\Phi \sim \Psi/r$, and the $m=0$ mode of $\Psi$ is approximately linear in $r$ near the origin.
At the outer radial boundary, $r=b L_r$, we expect radiative behavior, so we impose a 3rd-order radiation boundary condition~\cite{Osburn_2022,Bayliss1980}:
\begin{equation}
    (\partial_{r}^{3} - 3i\omega \partial_{r}^{2} - 3\omega^{2} \partial_{r} + i\omega^{3})\Psi\big|_{r=b L_r}=0 \, .
    \label{eq: bayliss-turkel}
\end{equation}
For $m=n=0$ the condition is a simpler regularity condition
\begin{equation*}
    \partial_{r} \Psi_{m=n=0}\big|_{r=b L_r}  = 0 \, .
\end{equation*}
At the polar boundaries, the full field must remain continuous and differentiable, satisfying the conditions
\begin{align}
&    \Psi_{m\ne 0}\big|_{\theta=0,\pi}=0 \, ,
    \label{eq:polar_{mn}_nonzero}
\\
& \partial_{\theta}\Psi_{m=0}\big|_{\theta=0,\pi}=0.
    \label{eq:polar_{mn}_zero}
\end{align}

In the numerical implementation, any derivatives that appear in boundary conditions are approximated using one-sided finite differences. Substitution of these finite-difference approximations into the boundary condition equations produces the corresponding boundary rows of the sparse matrix. Unlike the five-point stencil used at interior points, the boundary rows can involve several neighboring points because of the nature of the appropriate one-sided finite differences.

\begin{figure*}[t]
    \centering
    \includegraphics[trim={0 1 0 0},clip,width=\textwidth]{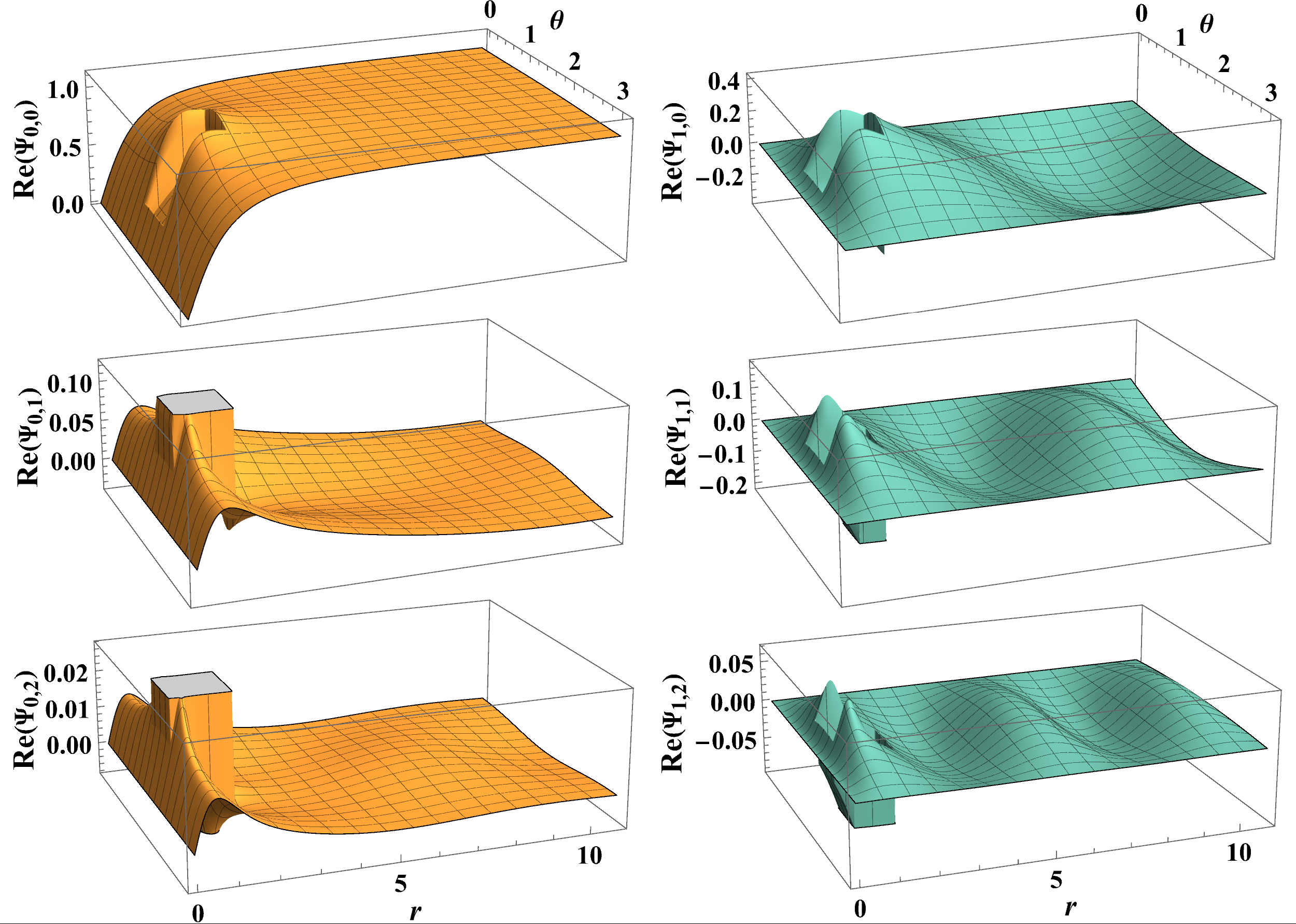}
    \caption{Numerical solutions, $\Psi_{mn}(r,\theta)$, of Eq.~\eqref{eq:matrix_equation} are shown for several modes with $e=0.3$, $\eta=0.5$, and $j=3$. The modes include $m=(0,1)$ and $n=(0,1,2)$. Our actual radial domain extends farther, but we truncate for illustrative purposes.  Notice the discontinuous jump from $\Psi^\text{ret}_{mn}$ to $\Psi^\mathcal{R}_{mn}$ across the worldtube. Two of the modes involve a positive jump to a larger $\Psi^\mathcal{R}_{mn}$ while the others involve a negative jump to a smaller $\Psi^\mathcal{R}_{mn}$. It is apparent how $\omega_{mn} = m\Omega_\phi+n\Omega_r$ corresponds to smaller wavelengths for higher mode numbers.
    }
    \label{fig:mn mode solution}
\end{figure*}

\section{Results and Analysis}
\label{sec:results}
To evaluate the effective source and puncture for a given $m$-mode we developed code in Mathematica. We use extended precision in Mathematica to handle catastrophic cancellations in the effective source expressions near the particle, although recent work has avoided these cancellations and the associated need for extended precision~\cite{vu2026selfforcecalculationsnumericalrelativity}. Unfortunately, we have not implemented that fix near the particle, so our extended precision code is slow: for each $m$ it requires approx. 1~hour to produce data for $j=2$ and approx. 20~hours to produce data for $j=3$ on a desktop workstation.

In contrast, numerically solving \eqref{eq:matrix_equation} in python takes roughly 30 seconds on a laptop after minor optimizations. Specifically, we solve using \texttt{pydiso}~\cite{pydiso}, a python wrapper of Intel MKL Pardiso~\cite{MKL}. Our linear systems are small enough that the Pardiso solver can handle them without encountering excessive speed or memory limitations.

\begin{figure*}[t]
    \centering
    \includegraphics[width=\textwidth]{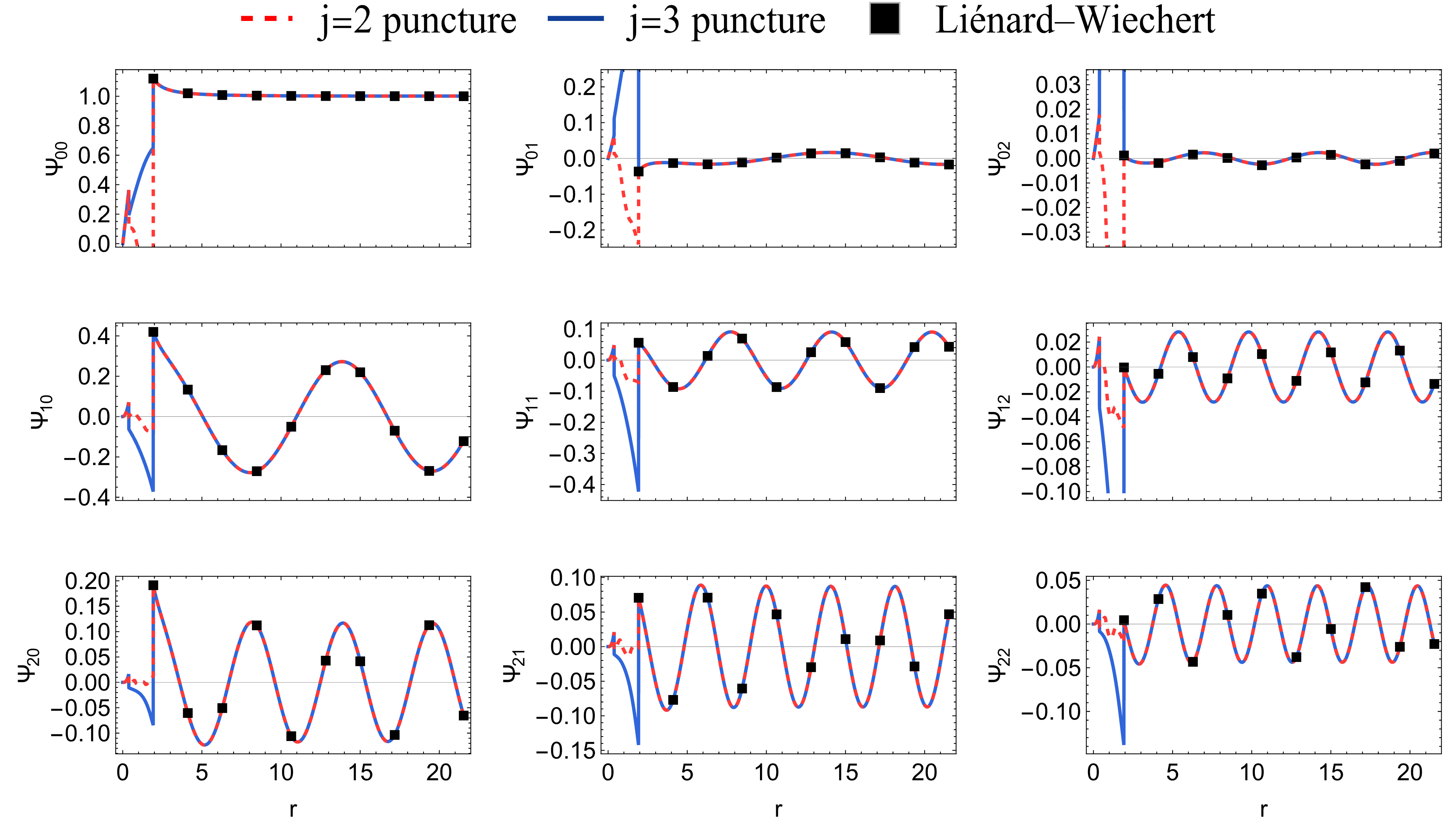}
    \caption{Comparison of our numerically calculated $\Psi^\text{ret}_{mn}$ to appropriate modes of the Li\'{e}nard--Wiechert potential for $e=0.3$ and $\eta=0.5$. We see that the $j=2$ and $j=3$ punctures+effective sources produce agreement with each other and agreement with the Li\'{e}nard--Wiechert potential over multiple wavelengths for $\theta=\pi/2-\Delta\theta/2$. Within the worldtube, the two punctures exhibit different behavior, which is expected~\cite{Wardell_2012}. The limiting factor of this comparison is resolution of the $t$ and $\phi$ integrals to extract each mode of the Li\'{e}nard--Wiechert potential, which produces an approximately $\sim 1\%$ relative error.}
    \label{fig:lw comp}
\end{figure*}

We show several $n$ modes of the numerical solution for $m=0$ and $m=1$ in Figure \ref{fig:mn mode solution}. Notice the expected discontinuities at the worldtube interfaces and distant wavelike behavior. Figure \ref{fig:lw comp} compares the numerical behavior of $\Psi^\text{ret}_{mn}$ to that of appropriate $\phi$ and $t$ integrals of the Li\'{e}nard--Wiechert potential (Eq.~\eqref{LW pot}), showing agreement. Sufficient $t,\phi$ sampling of the Li\'{e}nard--Wiechert potential required significant computation time, but the agreement between methods improved with increasing resolution in $t$ and $\phi$.

\begin{figure}
    \centering
        \includegraphics[width=\columnwidth]{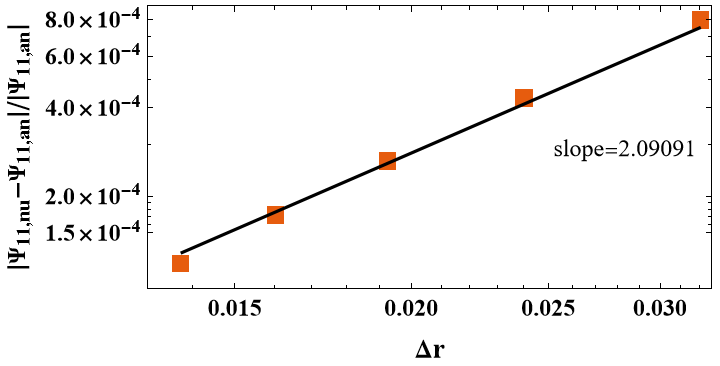}
    \hfill
        \includegraphics[width=\columnwidth]{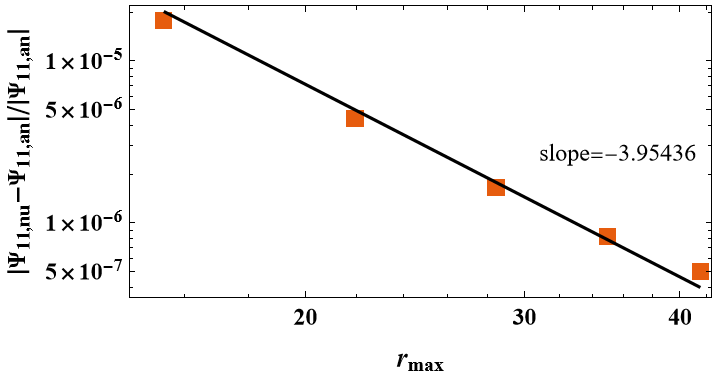}
    \caption{Measuring numerical convergence of our solution at $r/p=21.5$  for $e=0.3$ and $\eta=0.5$ with decreasing grid spacing (top), our second order scheme has error with a fitted order of 2.09 (with an expected order of 2). Similarly, our boundary scheme of expected order $r_{max}^{-4}$ produced a fitted order -3.95 for increasing outer radius (bottom).
    \label{fig:convergence}}
\end{figure}

Our ultimate goal is to measure convergence of the specific elements of our scheme that are not present for circular orbits, which will primarily involve quantifying convergence with $n$. However, before we analyze those aspects, we must validate our PDE solutions by demonstrating associated numerical convergence. Throughout this work, we employ a second order accurate finite difference method, where the step sizes in $r$ and $\theta$ are scaled proportionally such that $\Delta r/\Delta\theta  \sim 5 L_r/(2\pi)$. Thus, our numerical solution should have an error proportional to $\Delta r^{2}$, which Fig.~\ref{fig:convergence} verifies.\footnote{We only consider the $j=3$ case for the self-consistency tests because the non-smoothness of $j=2$ limits other aspects of convergence before resolution becomes a factor.}

We next verify that the outer boundary is far enough that our approximate radiative boundary condition is not a limiting factor. At the large $r$ boundary, we impose a condition with an error of size $\mathcal{O}(1/r^{4})$. Calculating the average error over one wavelength, we find that our numerical scheme matches our expectations, see Fig.~\ref{fig:convergence}. At the $\theta$ boundaries and $r=0$, the differential representation of our boundary conditions are exact. Thus, the finite difference representation of these boundaries does not contribute any error besides the discretization error that has already been examined.

\subsection{Mode convergence}
\label{sec: Convergence}
With sufficient assurance that we have achieved the expected numerical convergence, we advance to measuring mode convergence. First we will consider frequency mode convergence. To do this, we consider the behavior of the positive $n$-mode coefficients for the residual field, $\Psi^\mathcal{R}_{[j]mn}$, as a function of $n$ (negative $n$-modes converge similarly).

In the case of a perfectly smooth ($C^{\infty}$) function of time, we would expect to see exponential convergence in $n$ with higher eccentricities requiring more total $n$ modes. However, our time-domain functions are non-smooth at the particle, which will introduce the Gibbs phenomenon and should spoil exponential convergence at some level. Quantifying the size of this Gibbs obstacle is one of the main goals here. For non-smooth functions of time, we expect power-law convergence that improves with puncture order. Since the field equations involve second derivatives, the residual field will be two orders smoother than its corresponding effective source, and should therefore converge more rapidly. Therefore, quantities based on the third-order puncture should converge faster than those based on the second-order puncture. Because the non-smoothness is confined to the particle positions, we expect poorer convergence near the equatorial plane between $r_\text{max}$ and $r_\text{min}$. Discretizing $\theta$ with an odd number of spaces will avoid the exact position of non-smoothness, and it is not clear exactly how much difference there will be in convergence of the residual field at $\theta=\pi/2-\Delta\theta/2$ compared to the edge of the worldtube at $\theta=\pi/2-\pi/10$.

\begin{figure}
    \centering
    \includegraphics[width=\linewidth]{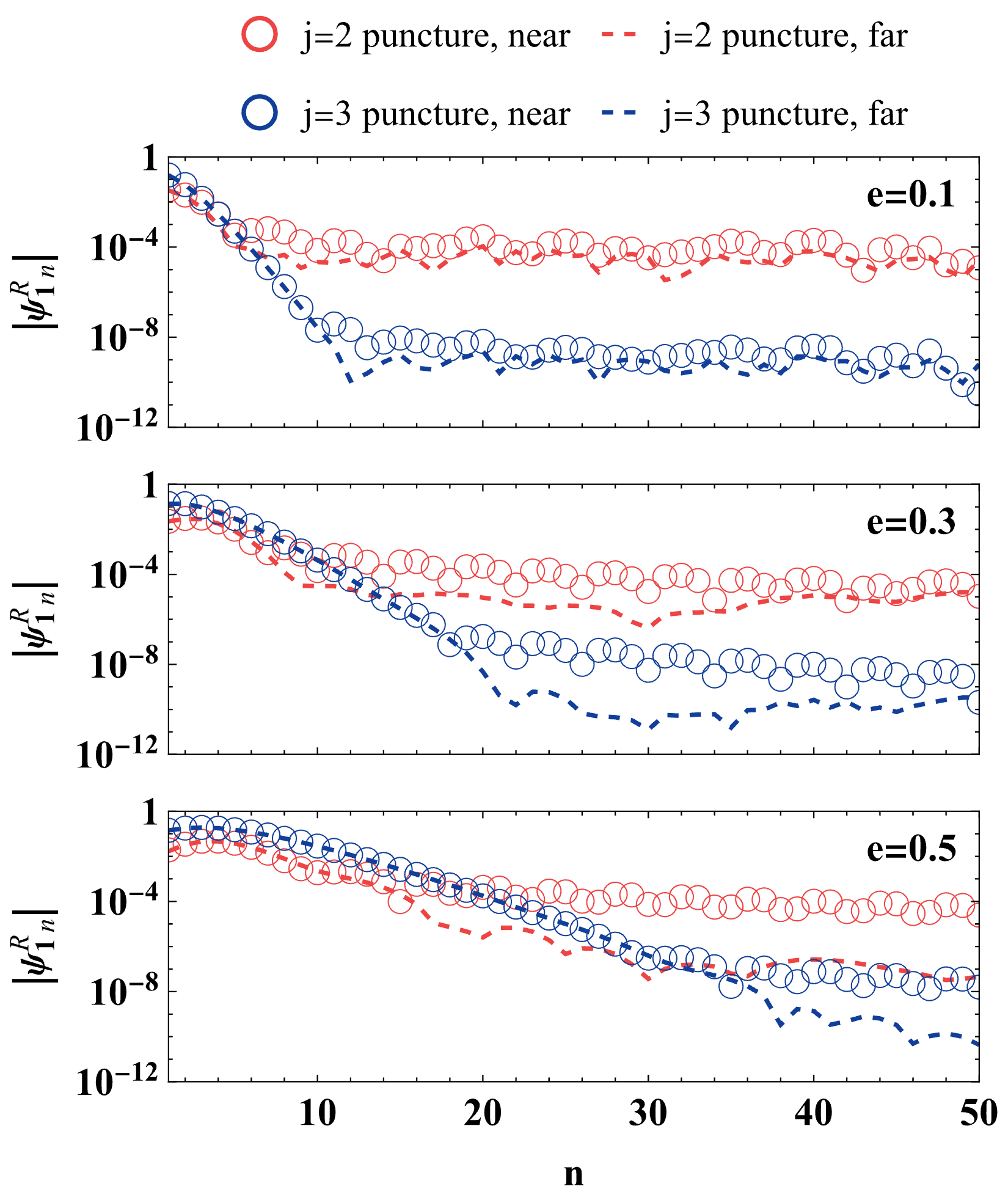}
    \caption{Convergence of the residual field with increasing $n$ for $m=1$, $r=r_p(T_r/8)$, $\eta=0.5$, $e=0.1$ (top), $e=0.3$ (middle), $e=0.5$ (bottom), $\theta=\pi/2-\Delta\theta/2$ (near) and $\theta=\pi/2-\pi/10$ (far). The Gibbs transition from exponential to power-law convergence is apparent when the behavior near the worldline deviates from the behavior far from the worldline. For $e=0.1$, the modes converge sufficiently fast to disguise the power-law transition; higher eccentricities involve a clear power-law transition that limits achievable accuracy. Increasing the puncture order maintains exponential convergence longer and achieves faster power law convergence. The $j=2$ noise floor is clearly too high for accurate self-force calculations, but $j=3$ enables frequency convergence to roughly 1 part in $10^7$.}
    \label{fig:reg field conv with n}
\end{figure}

\begin{figure}
    \centering
    \includegraphics[width=\linewidth]{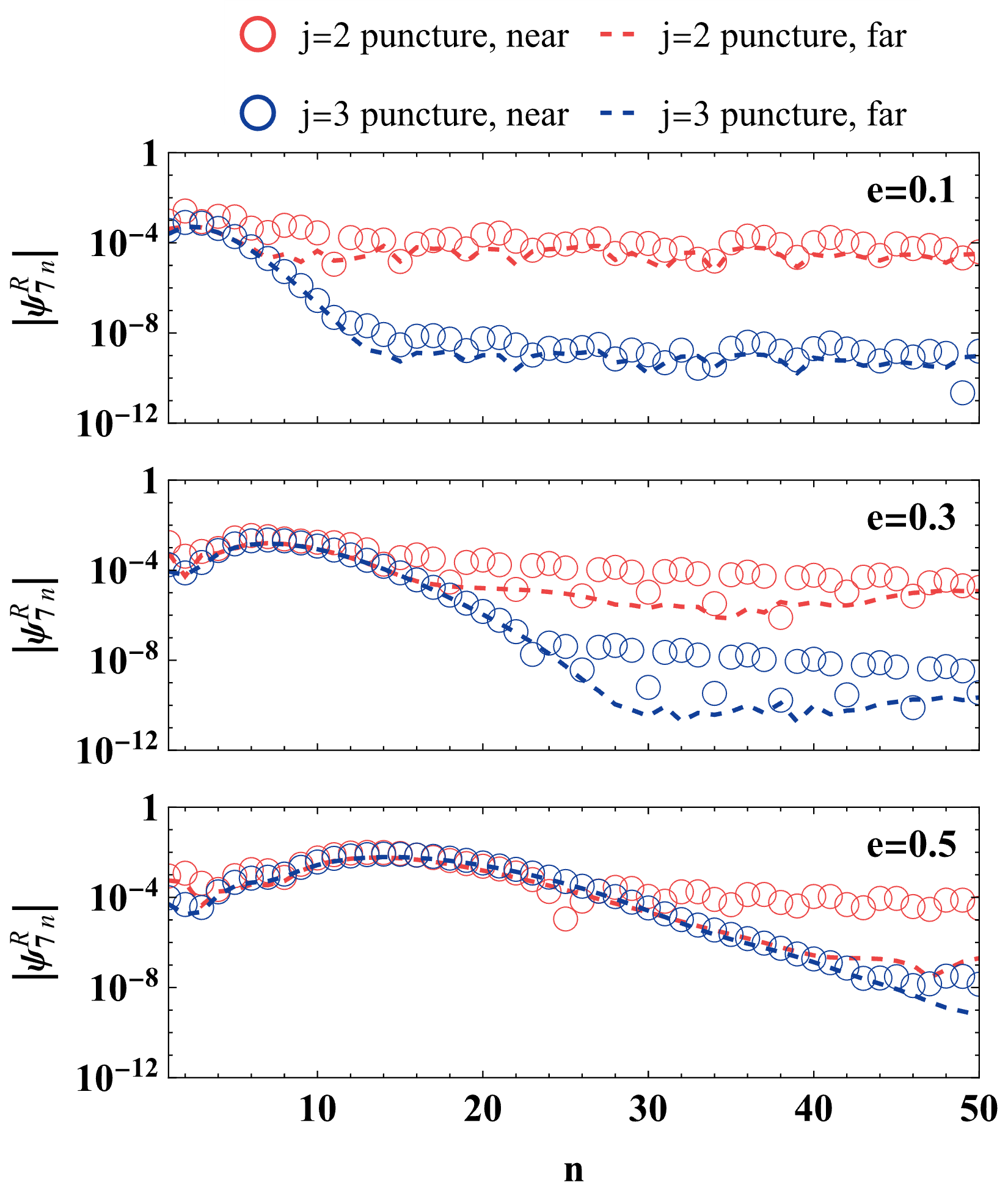}
    \caption{Convergence of the residual field with increasing $n$ for $m=7$, $r=r_p(T_r/8)$, $\eta=0.5$, $e=0.1$ (top), $e=0.3$ (middle), $e=0.5$ (bottom), $\theta=\pi/2-\Delta\theta/2$ (near) and $\theta=\pi/2-\pi/10$ (far). Features similar to the $m=1$ case of Fig.~\ref{fig:reg field conv with n} are present, but with the following additions: the largest amplitude mode is shifted to a higher $n$ for $m=7$, and that largest mode is not as big as for $m=1$. Similar power-law transitions that limit achievable accuracy are present.}
    \label{fig:reg field conv with n_7}
\end{figure}

Figures~\ref{fig:reg field conv with n} and~\ref{fig:reg field conv with n_7} demonstrate that the residual field converges more slowly with $n$ for cases closer to the worldline, with higher eccentricity, and/or with lower puncture order; these follow expectations. Additionally, the $j=3$ puncture produces relatively accurate behavior while the $j=2$ puncture introduces a high noise floor, which we believe is related to its blowup approaching the particle. For $e=0.1$, the modes converge sufficiently fast to disguise the power-law transition; higher eccentricities involve a clear power-law transition that limits achievable accuracy. Increasing the puncture order maintains exponential convergence longer and achieves faster power law convergence. To quantify the effect of the Gibbs phenomenon, observe how the $j=3$, $e=0.3$ case requires a minimum $|n_\text{max}| = 20$ to reveal the power-law transition at a relative truncation error of $\sim 10^{-6}$, after which increasing to $|n_\text{max}| = 50$ only improves the error by one additional order of magnitude. One interesting feature that was not necessarily expected is how the highest eccentricity has mostly converged before the Gibbs phenomenon causes a problematic power-law transition; we speculate further about this feature in Sec.~\ref{sec:conclusions}.


\begin{figure}
    \centering
    \includegraphics[width=\linewidth]{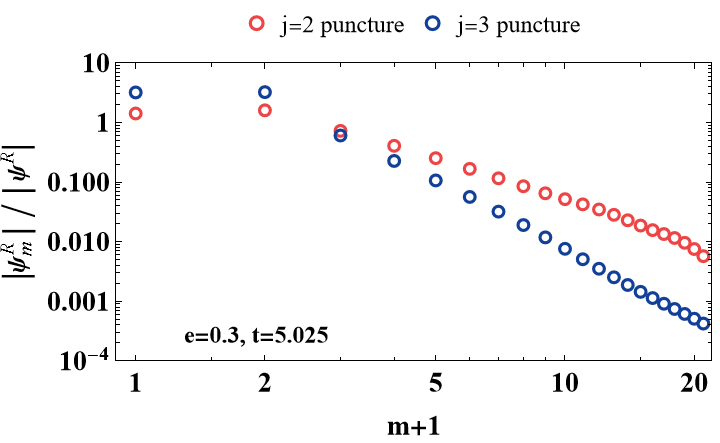}
    \caption{Convergence of the residual field $m$-modes for $e=0.3$ and $\eta=0.5$ at $t/p=5.025$ with $j=2$ and $j=3$ punctures. By $m=20$, the third order puncture has converged to approximately $10^{-3}$ and seems to have a steeper slope compared to the $j=2$ puncture, which clearly has not converged. Notice that the $j=2$ puncture involves artifacts at high $m$, where a sudden unphysical downturn near $m=17$ is present; issues like this seem consistent with the overall poor numerical behavior of the $j=2$ puncture.}
    \label{fig:m convergence}
\end{figure}

Now consider convergence with $m$. Figure~\ref{fig:m convergence} displays the relative magnitude of the residual field, $|\Psi^\mathcal{R}_{[j]m}|$, as a function of $m$ for $e=0.3$. Note that the sum over $n$-modes has been performed at a fixed time $t/p\simeq 5.025$ and for a position on the worldline. This dimension of convergence is not expected to be harmed by eccentricity, so we impose a moderate ceiling of $|m_\text{max}|=20$ rather than chasing comprehensive convergence; this $m$ ceiling turns out to be the overall limiting factor for convergence of the residual field, introducing an average relative error of $10^{-3}$ for the $j=3$ puncture. Notice that the $j=2$ puncture has clearly not converged, and also has artifacts at high $m$, see Fig.~\ref{fig:m convergence}. If this dimension of convergence were a priority, we could have investigated techniques such as $m$-mode regularization parameters, but we leave such extensions for future work.

\subsection{Reconstructed regular field}
\label{sec: Comparison to Regular Field}

Equipped with numerical solutions across $m$ and $n$, we now compare the total numerical residual field at the particle to the analytical regular field of  Eq.~\eqref{analytic potential}, which was obtained from the local expansions of the singular and regular fields described in Sec.~\ref{sec:analytic_regular_field}; recall that the availability of an exact comparison was one of the advantages of this toy model.

\begin{figure}
    \centering
    \includegraphics[width=\linewidth]{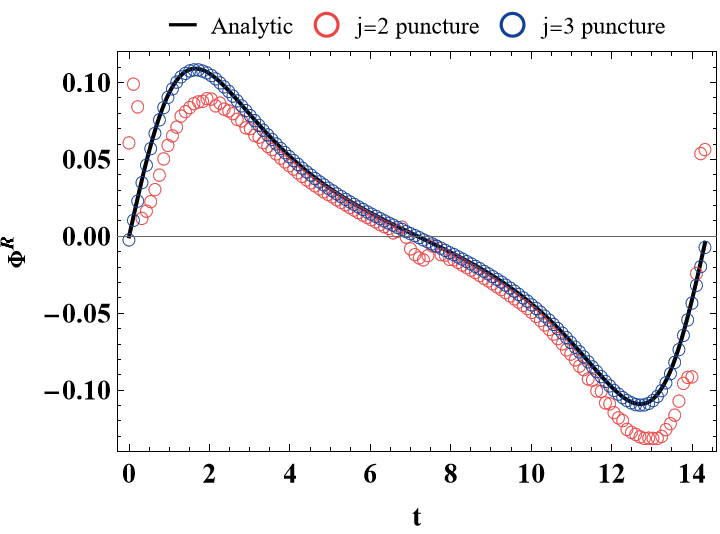}
    \caption{The total regular field at the charge position is shown for an orbit with $e=0.3$ and $\eta=0.5$ with the $j=2$ and $j=3$ puncture. The exact behavior is provided by the solid line. Interpolating our numerical solutions to the particle position, we see that with 41 total $m$-modes and 61 total $n$ modes (including positive and negative), the $j=3$ solution has converged quite well, while the $j=2$ solution still has visible discrepancies. Considering the prior analysis of $n$ and $m$ convergence for the $j=2$ case, such discrepancies are not surprising; nevertheless, it provides a valuable qualitative comparison to estimate the improvements introduced by increasing puncture order.
 }
    \label{fig:regular field agreement}
\end{figure}

The exact expression for $\Phi^R$ involves the velocity and acceleration of the moving charge as functions of $t$, which are straightforward to calculate numerically from the equations of motion (in the same way we did to evaluate the puncture and effective source). Then we compare Eq.~\eqref{analytic potential} to the numerical residual field for both the $j=2$ and $j=3$ punctures. Figure~\ref{fig:regular field agreement} demonstrates agreement of the numerical residual field from the $j=3$ puncture within the accuracy convergence implied by the sum over $m$-modes (see Fig.~\ref{fig:m convergence}). The $j=2$ puncture disagrees within an error range of $5\%$-$90\%$ error, which is not surprising based on its overall poor numerical behavior; nevertheless, it provides a valuable qualitative comparison to estimate the improvements introduced by increasing puncture order.

\section{Conclusions and Future Work}
\label{sec:conclusions}

We have performed numerical self-force style calculations of a particle's regular field during eccentric motion via the frequency-domain $m$-mode approach for the first time. Employing a fast Fourier transform allows us to determine all $n$ modes of the effective source in one efficient calculation, which significantly reduces the total computation time compared to mode-by-mode numerical integration. We analyze mode convergence of the solution at different puncture orders, $j=2$ and $j=3$, to estimate the degree of effective source smoothness necessary for accurate and efficient non-circular gravitational self-force calculations. We found the $j=2$ puncture to be too rough for accurate self-force results, but it nevertheless enabled an order-of-magnitude comparison of frequency convergence rates between $j=2$ and the significantly more accurate (and significantly more complicated) $j=3$ puncture. Based on this analysis, $j=3$ is unlikely to be a sufficiently high puncture order for moderate eccentricities ($e\simeq0.3$), but the considerable improvement from $j=2$ to $j=3$ suggests a $j=4$ or $j=5$ puncture could enable accurate gravitational self-force calculations for these eccentricities. 

A perhaps unexpected observation involves $n$-mode convergence for the highest eccentricity case we considered, $e=0.5$. For this case, the Gibbs phenomenon transition to slower power-law convergence was not visible until the solution had almost entirely converged. We speculate that this is a reflection of how even perfectly smooth highly eccentric sources produce a wide plateau of $n$-modes before the eventual transition to ideal exponential convergence as a large-$n$ tail; because even the best achievable behavior involves an increasingly delayed onset of true exponential convergence as $e$ grows, there seems to be more parity between smooth (far from the particle) and $j=3$ non-smooth (near the particle) behavior up until an ideal high-eccentricity exponential tail transition would occur. Arguably, this implies that the preexisting eccentricity issues present across even ideal frequency-domain scenarios are already a limiting factor, which lessens the relative severity of issues introduced by non-smooth time-dependence of the effective source for high-eccentricity cases (although additional investigation is warranted). 

These elliptic PDE results are based on a somewhat simple second-order finite difference scheme, which was chosen for rapid prototyping and testing purposes. When advancing to the gravitational case it will be vital to improve computational performance because the symmetric tensorial nature of the gravitational fields increases the number of unknowns by a factor of 10 (for direct metric perturbation approaches~\cite{osburn2026schwarzschildperturbationslorenzgauge,nishimura2026}). More sophisticated techniques, such as spectral discontinuous Galerkin (DG) methods, adaptive mesh refinement, matrix preconditioning, and hyperboloidal compactification~\cite{Zengino_lu_2008,Zengino_lu_2011a,Zengino_lu_2011b}, would enable significantly faster calculations and improved accuracy; those techniques are already available within the SpECTRE numerical relativity code~\cite{spectrecode}, which is a natural fit for future non-circular gravitational self-force calculations via elliptic PDEs, which are in development~\cite{Kelly2026}.

Despite having 2D PDEs as the focus of this work, the aforementioned SpECTRE's elliptic solver was originally designed for 3D PDEs (for numerical relativity initial data~\cite{Vu_2022,Vu_2024}). Future work could reintroduce $\phi$ as a variable but still solve in the frequency domain, making the field equations a set of 3D elliptic PDEs. This could allow researchers to more naturally interface with the capabilities of SpECTRE and might even simplify the effective source method; here we found that the need for analytic $m$-mode integrals of the puncture drastically complicates the effective source calculation. In contrast, a formulation of the PDEs in 3D would not require any analytic puncture integrals at all; this could be especially fortunate considering that there have not been any explicit gravitational $m$-mode puncture and effective source calculations for generic Kerr orbits, and that challenge would be significantly easier if the $m$-mode integrals were optional. Thus, a shift to 3D PDEs may simultaneously alleviate the burden of puncture calculations and streamline SpECTRE integration.

Another valuable extension would involve improving the mediocre rate of frequency convergence due to our non-smooth moving source and the associated Gibbs phenomenon inflicting slower power-law convergence. Higher-order punctures would alleviate this obstacle somewhat, but would fall short of true exponential convergence. It should be noted that the full separation of variables approach has largely overcome the Gibbs phenomenon through techniques such as the method of extended homogeneous solutions~\cite{Barack_2008} and the method of extended particular solutions~\cite{Hopper_2013} (although, even those ideal scenarios face high-eccentricity limitations). However, it is not obvious how (or whether) these techniques could generalize to generic Kerr orbits for the frequency-domain $m$-mode approach, although certain clever strategies may justify further investigation~\cite{Leather_2023,Whittall_2025}. Nevertheless, there is a guaranteed way to restore exponential frequency convergence: transform to a dynamical numerical domain in which the source position is fixed~\cite{Vaswani_2026,Field_2009}. Such a co-moving transformation would enable perfectly smooth time dependence by confining the effective source non-smoothness to spatial directions. But there is a catch: this time-dependent coordinate transformation would defeat separability of the time variable. Therefore, it would be important to determine whether the advantage of exponential frequency convergence justifies the penalty of coupling all the frequency modes together in one large system.

Finally, after future extensions of this work produce first-order Kerr metric perturbations for non-circular orbits, it will unlock the original purpose of the frequency-domain $m$-mode strategy: source construction for second-order Kerr metric perturbations. Such second-order Kerr calculations are in progress for circular motion~\cite{Leather:2026zhl,nishimura2026,leather2026b}. Constructing a non-circular second-order source by appropriately convolving first-order Fourier modes should lead to corresponding second-order solves that determine non-circular second-order self-force effects and their impact on waveform predictions. 

\acknowledgements
We thank Cillian Kelly, Jonathan Thompson, Nils Vu, Barry Wardell, Gavin George, Nami Nishimura, Charles Evans, Adam Pound, Ayush Roy, Rodrigo Panosso Macedo, Benjamin Leather, Christopher Whittall, Aditya Vaswani, Nils Deppe, Robbie Sinesi, Fintan Hughes, and Eli Hauber for helpful discussions. ZB, CR, and TO gratefully acknowledge support from the National Science Foundation under Grant No. PHY-2309020. ZB would like to thank the Geneseo foundation for their support through the Summer Research Fellowship and Wendell and Barbara Rhodes award. ZB acknowledges support from the NASA New York Space Grant. TO gratefully acknowledges support from the Fulbright U.S. Scholar Program, which is sponsored by the U.S. Department of State and the Fulbright Commission in Ireland; the contents are solely the responsibility of the authors and do not necessarily represent the official views of the Fulbright Program, the U.S. Government, or the Fulbright Commission in Ireland. JB acknowledges support from the National Science Foundation under Grants No.~PHY-2407742; No.~PHY-2607486; No.~OAC-2513338; and NASA award No.~80NSSC26K0340 at Cornell. JB also acknowledges support by the Sherman Fairchild Foundation at Cornell. LRV gratefully acknowledges support from the Ronald E. McNair Postbaccalaureate Achievement Program.

\appendix

\onecolumngrid
\section{Puncture expansion coefficients}
\label{app:coeffs}

The following provide full expressions for the singular field coefficients of Eq.~\eqref{phi of As}
\begin{align}
&A_{[1]}= \frac{q}{\sqrt{1-v^2+\big(\hat{\rcurs}\cdot\vec{v}\big)^2}}\, ,
\\
& A_{[2]} = -\frac{q}{2}\Bigg( \frac{\hat{\rcurs}\cdot\vec{a}}{\left(1-v^2+(\hat{\rcurs}\cdot\vec{v})^2 \right)^{3/2}} + \frac{ (\vec{v}\cdot\vec{a})(\hat{\rcurs}\cdot\vec{v})\left( 3-3v^2+2(\hat{\rcurs}\cdot\vec{v})^2 \right)}{(1-v^2)^2\left( 1-v^2+(\hat{\rcurs}\cdot\vec{v})^2 \right)^{3/2}} \Bigg) ,  
\\
& A_{[3]} = \frac{q}{2} \Bigg(  
\frac{3(\vec{v}\cdot\vec{a})^2\left(16(\hat{\rcurs}\cdot\vec{v})^6 +  40(1-v^2)(\hat{\rcurs}\cdot\vec{v})^4 + 30(1-v^2)^2(\hat{\rcurs}\cdot\vec{v})^2+5(1-v^2)^3 \right)}{4(1-v^2)^4\left(1-v^2+(\hat{\rcurs}\cdot\vec{v})^2\right)^{5/2}}
\\&\qquad\qquad + \frac{ (\hat{\rcurs}\cdot\dot{\vec{a}})(\hat{\rcurs}\cdot\vec{v})\left( 3-3v^2+2(\hat{\rcurs}\cdot\vec{v})^2 \right)}{3(1-v^2)^2\left( 1-v^2+(\hat{\rcurs}\cdot\vec{v})^2 \right)^{3/2}}
    + \frac{\big(4(\vec{v}\cdot\dot{\vec{a}})+3a^2\big) \left( 8(\hat{\rcurs}\cdot\vec{v})^4 + 12(1-v^2)(\hat{\rcurs}\cdot\vec{v})^2+3(1-v^2)^2 \right)}{12(1-v^2)^3\left( 1-v^2+(\hat{\rcurs}\cdot\vec{v})^2 \right)^{3/2}} \notag
\\&\qquad\qquad + \frac{3(\hat{\rcurs}\cdot\vec{a})^2}{4\left(1-v^2+(\hat{\rcurs}\cdot\vec{v})^2\right)^{5/2}} +\frac{(\hat{\rcurs}\cdot\vec{a})(\vec{v}\cdot\vec{a})(\hat{\rcurs}\cdot\vec{v})\left( 8(\hat{\rcurs}\cdot\vec{v})^4+20(1-v^2)(\hat{\rcurs}\cdot\vec{v})^2+15(1-v^2)^2 \right)}{2(1-v^2)^3\left( 1-v^2+ (\hat{\rcurs}\cdot\vec{v})^2 \right)^{5/2}}  \Bigg) , \notag
\end{align}
where we have dropped the $p$ subscripts such that $\vec{v}_p=\vec{v}$,  $\vec{a}_p=\vec{a}$, and  $\dot{\vec{a}}_p=\dot{\vec{a}}$ for brevity.
\twocolumngrid

\bibliography{main}

\end{document}

%% file: figures/lc-static-t0-down.tikz
\begin{tikzpicture}[
  >=Latex,
  cone/.style={thick},
  wl/.style={thick},
  vec/.style={->, thick},
  event/.style={circle, fill=black, inner sep=1.3pt},
  lab/.style={font=\small}
  ]

  \def\L{3.0}
  \def\R{1.5}

  \coordinate (X) at (0,0);
  \node[event,label={[lab]right:$x$}] at (X) {};

  \draw[cone] (X) -- ++(-\L, \L);
  \draw[cone] (X) -- ++(-\L,-\L);
  \draw[cone] (X) -- ++( \R, \R);
  \draw[cone] (X) -- ++( \R,-\R);

  \def\xw{-2.2}
  \coordinate (Wb) at (\xw,-3.6);
  \coordinate (Wt) at (\xw, 3.6);
  \draw[wl] (Wb) -- (Wt);
  \node[lab] at (\xw+0.7,-2.8) {$x_p(\tau)$};

  \coordinate (T0) at (\xw,-0.85);
  \node[event,label={[lab]left:$x_p(\tau_0)$}] at (T0) {};

  \draw[vec] (T0) -- (X) node[midway,above,lab] {$\bar\gamma^\alpha$};

  \coordinate (Zplus)  at (\xw, {abs(\xw)});
  \coordinate (Zminus) at (\xw, {-abs(\xw)});

  \node[event,label={[lab]left:$x_p(\tau_+)$}] at (Zplus) {};
  \node[event,label={[lab]left:$x_p(\tau_-)$}] at (Zminus) {};

  \draw[
  decorate,
  decoration={brace, amplitude=4pt}
  ] (T0) -- (Zplus)
  node[midway,left=6pt,lab] {$s_+$};

  \draw[
  decorate,
  decoration={brace, amplitude=4pt}
  ] (Zminus) -- (T0)
  node[midway,left=6pt,lab] {$-s_-$};

\end{tikzpicture}